\documentclass[a4paper,journal,11pt,draftclsnofoot,onecolumn]{IEEEtran}

\usepackage{dsfont}
\usepackage{amsmath}
\usepackage{amssymb}
\usepackage{amsfonts}
\usepackage{graphicx}
\usepackage{amsmath}
\usepackage[all,poly]{xy}
\usepackage{multirow}
\usepackage{algorithm}
\usepackage{algorithmic}
\usepackage{color}
\usepackage[noadjust]{cite}
\usepackage{caption}
\usepackage{subfigure}
\usepackage{url}

\usepackage{tikz,pgf}
\usetikzlibrary{fit}
\usetikzlibrary{arrows,automata}
\usepackage{verbatim}
\usetikzlibrary{positioning,shapes.geometric}
\usepackage{url}
\usepackage[T1]{fontenc}

\newcommand{\figwidth}{\columnwidth}
\newcommand{\bone}{\boldsymbol{1}}

\newcommand{\bthe}{\boldsymbol{\theta}}
\newcommand{\bSig}{\boldsymbol{\Sigma}}

\newcommand{\bOme}{\boldsymbol{\Omega}}
\newcommand{\bPsi}{\boldsymbol{\Psi}}
\newcommand{\bLam}{\boldsymbol{\Lambda}}

\def\bsa{{\boldsymbol{a}}}

\def\bsc{{\boldsymbol{c}}}

\def\bse{{\boldsymbol{e}}}

\def\bsm{{\boldsymbol{m}}}

\def\bsp{{\boldsymbol{p}}}
\def\bsq{{\boldsymbol{q}}}

\def\bss{{\boldsymbol{s}}}
\def\bst{{\boldsymbol{t}}}

\def\bsx{{\boldsymbol{x}}}
\def\bsy{{\boldsymbol{y}}}
\def\bsz{{\boldsymbol{z}}}

\def\bsA{{\boldsymbol{A}}}

\def\bsC{{\boldsymbol{C}}}

\def\bsK{{\boldsymbol{K}}}

\def\bsM{{\boldsymbol{M}}}

\def\bsS{{\boldsymbol{S}}}
\def\bsT{{\boldsymbol{T}}}

\def\bsY{{\boldsymbol{Y}}}

\def\dsR{{\mathds{R}}}

\def\calS{{\mathcal{S}}}

\def\calB{{\mathcal{B}}}

\def\calN{{\mathcal{N}}}

\def\calS{{\mathcal{S}}}

\newcounter{algo}
\renewcommand{\thealgo}{\arabic{algo}}

\title{Unsupervised Unmixing of Hyperspectral Images Accounting for Endmember Variability}

\author{\vspace{1cm}Abderrahim Halimi, Nicolas Dobigeon and Jean-Yves Tourneret\\
\vspace{0.5cm}
\normalsize University of Toulouse, IRIT/INP-ENSEEIHT/T\'eSA, Toulouse, France\\
\small\texttt{\{Abderrahim.Halimi,Nicolas.Dobigeon,Jean-Yves.Tourneret\}@enseeiht.fr}\\
\thanks{Part of this work has been funded by the Hypanema ANR Project n$^\circ$ANR-12-BS03-003.}}

\begin{document}

\maketitle

\begin{abstract}
This paper presents an unsupervised Bayesian algorithm for hyperspectral image unmixing accounting for endmember variability. The pixels are modeled by a linear combination of endmembers weighted by their corresponding abundances. However, the endmembers are assumed random to take into account their variability in the image. An additive noise is also considered in the proposed model generalizing the normal compositional model. The proposed algorithm exploits the whole image to provide spectral and spatial information. It estimates both the mean and the covariance matrix of each endmember in the image. This allows  the behavior of each material to be analyzed and its variability to be quantified in the scene. A spatial segmentation is also obtained based on the estimated abundances. In order to estimate the parameters associated with the proposed Bayesian model, we propose to use a Hamiltonian Monte Carlo algorithm. The performance of the resulting unmixing strategy is evaluated via simulations conducted on both synthetic and real data.
\end{abstract}

%\begin{keywords}
%Hyperspectral imagery, endmember variability, image classification, spectral unmixing, Bayesian algorithm,  Hamiltonian Monte-Carlo, MCMC methods.
%\end{keywords}

%%%%%%%%%%%%%%%%%%%%%%%%%%%%%%%%%%%%%%%%%%%%%%%%%%%

\section{Introduction} \label{sec:Introduction}
Hyperspectral imaging is a remote sensing technology that collects $3$ dimensional data cubes composed of $2$D spatial images acquired in numerous contiguous spectra bands. Due to the limited spatial resolution of the observed image, each pixel generally consists of several physical elements that are linearly \cite{Keshava2002,Heinz2001}  or nonlinearly \cite{Bioucas2012,Dobigeon2014,Halimi2011TGRS} mixed. Spectral unmixing (SU) consists of decomposing the pixel spectrum to recover these materials, known as \emph{endmembers}, and estimating the corresponding proportions or \emph{abundances} \cite{Dobigeon2009}. The linear mixture model (LMM) has received great interest in the literature and has been used intensively for SU. The unmixing is generally performed using two distinct steps: (i) identifying the endmembers using an endmember extraction algorithm (EEA) such as vertex component analysis (VCA) \cite{Nascimento2005}, pixel purity index (PPI) \cite{Boardman1993} and N-FINDR \cite{Winter1999}, (ii) estimating the abundances under physical non-negativity and sum-to-one constraints using algorithms such as the fully constrained least squares \cite{Heinz2001}. Some algorithms also tackle the SU problem in an unsupervised manner, i.e., by jointly estimating the endmembers and the abundances. This is generally achieved under a statistical framework using optimization techniques \cite{Nascimento2012} or Markov chain Monte Carlo (MCMC) simulation methods \cite{Dobigeon2009,Altmann2014b}. The unsupervised algorithms generally provide more sophisticated results and appear to be less sensitive to the absence of pure pixels \cite{Bioucas2012}.

\begin{figure}[h!]
\centering \subfigure[]{\includegraphics[width=0.32\figwidth]{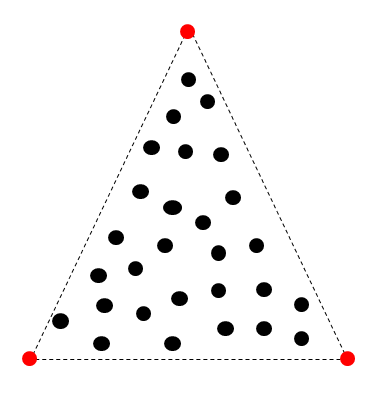}}
\subfigure[]{\includegraphics[width=0.32\figwidth]{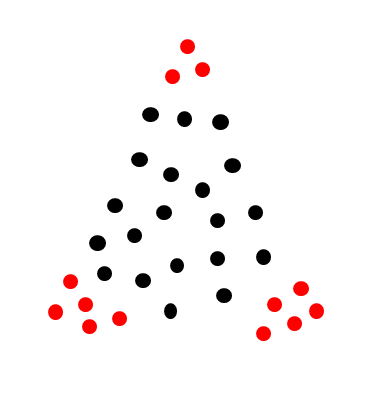}}
\subfigure[]{\includegraphics[width=0.32\figwidth]{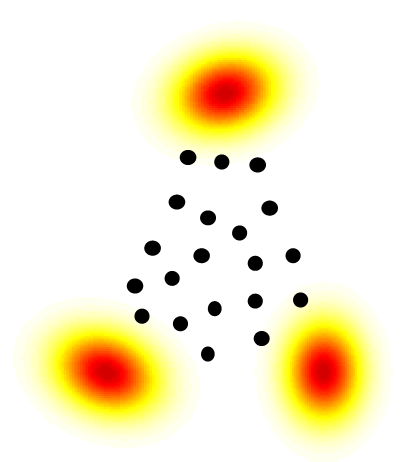}}
\caption{Simplex representation for  (a) endmembers without variability, (b) endmembers as a finite set (or bundle) and (c) endmembers as a distribution.  } \label{fig:Simplex}
\end{figure}

The previous described algorithms provide one endmember spectrum for each physical component present in the image (see Fig. \ref{fig:Simplex}(a)). This appears as a clear simplification since in many cases, the endmember spectra vary along the image causing what is known as \emph{spectral variability}. Spectral variability has been identified as one of the most profound sources of error in abundance estimation and is knowing growing interest in the hyperspectral community \cite{Somers2011,Zare2014}. Many algorithms have been proposed in the literature to describe this variability by considering each endmember as a finite set or as a statistical distribution. Some deterministic approaches represent each physical material as a set or bundle of spectra (see Fig. \ref{fig:Simplex}(b)). One can distinguish between algorithms assuming a known spectral library \cite{Roberts1998,Bateson2000} and those estimating it from the data \cite{Goenaga2013,Somers2012}. SU resulting from these approaches is generally sensitive to the quality of the  available or extracted endmember libraries. There are also statistical approaches assuming that each endmember is a random vector with a given distribution  (see Fig. \ref{fig:Simplex}(c)). Statistical approaches provide a parametric representation of the endmembers and thus can estimate endmembers that are not present in the observed data. This property makes these algorithms more robust in absence of pure pixels \cite{Eches2010ip,Zare_conf2013,Zare2013}. A more detailed discussion about these algorithms, their advantages and challenges is available in \cite{Somers2011,Zare2014}.

The main contribution of this paper is the consideration of endmember variability under a Bayesian framework. Any endmember is considered via a probability distribution to model its variability. Two main approaches have been considered in the literature assuming the endmembers are random vectors: the Beta compositional model \cite{Zare_conf2013} and the normal compositional model (NCM) \cite{Stein2003,Eches2010ip,Zare2013}. This paper considers a generalization of the NCM model characterized by  Gaussian variability for the endmembers (as for the NCM) and an additive Gaussian noise modeling fitting errors (which was not present in the NCM). Moreover, the proposed model considers a different mean and covariance matrix for each endmember to analyze each component separately. These parameters are both estimated to generalize the works of \cite{Eches2010ip} and \cite{Zare2013} that estimated the endmembers means and covariances, respectively. Moreover, the endmember fluctuation with
respect to the spectral bands is quantified by considering non identically distributed endmember variances.

Another important point concerning hyperspectral unmixing is the spatial correlation between pixels. Indeed, even if many  algorithms consider a pixel-by-pixel context, recent studies have shown the interest of considering spatial information to improve the unmixing quality \cite{Bali2008,Eches2010tgrs,Chen2014}. Within a Bayesian framework, this spatial correlation can be introduced using Markov random fields (MRFs) as already shown in \cite{Bali2008,Eches2010tgrs,Rand2003}. In this work, a Potts model is considered since it has already shown good performance when processing hyperspectral images \cite{Bali2008,Eches2010tgrs}.
The image is then segmented into regions sharing similar abundance characteristics. Note that this segmentation was also achieved in \cite{Eches2010tgrs} and \cite{Nascimento2012} by considering Gaussian and Dirichlet distributions for the abundances.

This paper proposes an unsupervised Bayesian algorithm to estimate the parameters associated with endmembers and abundances. In addition to the abundance Dirichlet priors, it assumes appropriate prior for the remaining parameters/hyperparameters to satisfy the known physical constraints. The joint posterior distribution of the proposed Bayesian model is then derived. However, the classical minimum
mean square error (MMSE) and maximum a posteriori (MAP)
estimators cannot be easily computed from this joint posterior. A classical way of alleviating this problem is to generate
samples distributed according to the posterior using MCMC methods. This goal is achieved in this paper using a Gibbs sampler coupled with a Hamiltonian Monte Carlo (HMC) method. HMC is well adapted for problems with a large number of parameters to be estimated  \cite{Brooks2011}. Moreover, this method presents good mixing properties when compared to the classical Metropolis-Hasting algorithm. This paper considers a constrained-HMC (CHMC) that has been introduced in \cite[Chap.~5]{Brooks2011} and successfully used for hyperspectral SU in \cite{Altmann2014b}. This CHMC accounts for inequality constraints which is required to satisfy the physical constraints related to the proposed SU problem.

%Spectral unmixing
%endmember variability and image segmentation
%
%Contrib. 1: EV
%Contrib. 2: segmentation
%
%Bayesian algo.

% Spectres synth def.

The paper is structured as follows. The unmixing problem considered in this study is formulated in Section \ref{sec:Problem_formulation}. The different components of the proposed Bayesian model are studied in Section \ref{sec:Hierarchical_Bayesian_model}. Section \ref{sec:Hybrid_Gibbs_algorithm} introduces the Gibbs sampler and the CHMC method which will be used to generate samples asymptotically distributed according to the joint posterior of the unknown parameters and hyperparameters. Section \ref{sec:Simulation_results_on_synthetic_data} analyzes the performance of the proposed algorithm when applied to synthetic images. Results on real hyperspectral images are presented in Section \ref{sec:Simulation_results_on_real_data} whereas   conclusions and future works are reported in Section
\ref{sec:Conclusions}.

%%%%%%%%%%%%%%%%%%%%%%%%%%%%%%%%%%%%%%%%%%%%%%%%%%%
%%%%%%%%%%%%%%%%%%%%%%%%%%%%%%%%%%%%%%%%%%%%%%%%%%%
%%%%%%%%%%%%%%%%%%%%%%%%%%%%%%%%%%%%%%%%%%%%%%%%%%%
%LMM
\section{Problem formulation}\label{sec:Problem_formulation}
\subsection{Notations}
The variables used in this paper are described as follows

\begin{tabular}{ll}
$N$        &  number of pixels \\
$R$        &  number of endmembers \\
$L$        &  number of spectral bands \\
$K$        &  number of spatial classes \\
$\bsY \in \dsR^{L \times N}$   &  spectra of the pixels   \\
$\bsA \in \dsR^{R \times N }$  &  abundance matrix  \\
$\bsT \in \dsR^{R-1 \times N }$ &  reparameterized abundance matrix   \\
$\bsM \in \dsR^{L \times R}$      &  endmember means    \\
$\bSig \in \dsR^{R \times L}$  &  matrix containing the diagonal of endmember covariances \\
$\bPsi \in \dsR^{1 \times N }$ &  noise variances \\
$\bsK \in \dsR^{L \times N }$  &  matrix whose rows equals  $\bPsi$  \\
$\bsz \in \dsR^{1 \times N }$  & labels  \\
$\bsC\in \dsR^{R \times K }$   & Dirichlet parameters   \\	
%$\bOme = \bSig^T \left(A \odot  A\right) + \bsK$:  Global variance   (L X N2) \\
%$\bLam  = \frac{1}{\bOme}$: (L X N2) \\
\end{tabular}

\subsection{Mixing model and endmember variability} \label{subsec:Mixture_Models}
This section introduces the proposed mixture model. The classical LMM assumes the pixel spectrum $\bsy_n$, $n\in \left\lbrace 1,\cdots,N \right\rbrace$, where $N$ is the number of pixels in the image, is a linear combination of $R$ deterministic endmembers $\bss_{r}$, $r\in \left\lbrace 1,\cdots,R \right\rbrace$,   corrupted by an additive noise as follows
\begin{equation}
\bsy_n = \sum_{r=1}^{R}{a_{rn} \bss_{r}}+ \bse_{n} = \bsS \bsa_n + \bse_{n} \label{eqt:linear_model1}
\end{equation}
with
\begin{equation}
\bse_{n}    \sim \calN \left(\boldsymbol{0}_L, \psi_n^2 \mathbf{I}_L \right)
%\calN \left( 0, \kappa^2_\ell  \right), ~~ \ell = 1,\cdots,L
\label{eqt:linear_model2}
\end{equation}
where  $R$ is the number of endmembers, $\bsy_n$  is an $(L \times 1)$ vector representing the $n$th observed pixel, $L$ is the number of spectral bands,  $\boldsymbol{0}_L$ is an  ($L \times 1$) vector of $0$, $\mathbf{I}_L$ is the ($L \times L$) identity matrix, $\bsa_n = \left[a_{1n}, \cdots, a_{Rn} \right]^T$ is the $(R \times 1)$ abundance vector of the $n$th pixel, $\bsS=\left[\bss_{1}, \cdots, \bss_{R}\right]$ is an $(L \times R)$ matrix of endmembers and $\bse_{n}$ is a centered additive, independent and identically distributed Gaussian noise.

The endmembers are generally variable in the observed image due to environmental conditions or inherent variability \cite{Zare2014}. In this paper, we introduce a model taking into account this variability. The proposed model can be seen as a generalization of the NCM model (GNCM) since it introduces an additional residual Gaussian noise $\bse$ as follows
\begin{equation}
\bsy_n = \sum_{r=1}^{R}{a_{rn} \bss_{rn}}  + \bse_n = \bsS_n \bsa_n + \bse_n
 \label{eqt:Normal_compositional_model1}
\end{equation}
with
\begin{equation}
\bss_{rn}   \sim \calN \left(\bsm_{r} , \textrm{diag}\left( \boldsymbol{\sigma}^2_r\right)  \right)
\label{eqt:Double_convolution2}
\end{equation}
%\begin{eqnarray}
%\left\lbrace \begin{array}{l}
%\bss_{rn}   \sim \calN \left(\bsm_{r} , \textrm{diag}\left( \boldsymbol{\sigma}^2_r\right)  \right) \\
%\bse_n  \sim \calN \left(\boldsymbol{0}_L, \psi_n^2 \mathbf{I}_L \right),  %\psi^2_n << %\sigma^2_{r\ell}, \forall r, \forall \ell.
%\end{array} \right.
%\label{eqt:Double_convolution2}
%\end{eqnarray}
where  $\bsS_n  = \left[\bss_{1n},\cdots,\bss_{Rn}\right]$, $\boldsymbol{\sigma}^2_r = \left[ \sigma^2_{r1},\cdots,\sigma^2_{rL}\right]$ is the variance vector of the $r$th endmember and $\bsM = \left[\bsm_{1}, \cdots, \bsm_{R}\right]$ is the $(L \times R)$ matrix containing the endmember means of the image. The main difference between model \eqref{eqt:Normal_compositional_model1} and the LMM used in \cite{Dobigeon2009} is that the endmember matrix $\bsS_n$  depends on each observed pixel in order to introduce the spectral variability. Each physical element is then represented in a given pixel by an endmember $\bss_{rn} $ that has its own Gaussian distribution whose variances $\boldsymbol{\sigma}^2_r$ change from one band to another. This allows the GNCM to capture the spectral variations of each physical element with respect to each spectral band. The GNCM also includes an additional Gaussian noise $\bse_n \sim \calN \left(\boldsymbol{0}_L, \psi_n^2 \mathbf{I}_L \right)$ (that is independent from the variables $\bss_{1n},\cdots,\bss_{Rn}$)  whose goal is to make the proposed model more robust with respect to mismodeling. Moreover, we consider that the endmember variability is the main source of randomness in the observed pixel, which is ensured by assigning a very sparse prior to the noise variance (see Eq. \eqref{eqt:priorNoise}).
%which means that the noise variance $\psi^2_n $ is assumed to be smaller than the endmember variances $\sigma^2_{r\ell}, \forall r, \forall \ell$ (this hypothesis also avoids identifiability problems between the variances).
Note finally that the proposed model reduces to the NCM for
$\psi^2_n=0,~~ \forall n$. Thus, it generalizes the model of \cite{Eches2010ip,Zare2013} by considering a non-isotropic covariance matrix for each endmember.

%Indeed, releasing the constraint \ref{eqt:Double_convolution2} might result in biased estimates for the endmember variances as

\subsection{Abundance reparametrization} \label{subsec:Abundance_reparametrization}
Since the abundance vector $\bsa_n$   usually represents spatial coverage of the material in a given pixel, it should satisfy the physical positivity and sum-to-one constraints
\begin{equation}
a_{rn} \geq 0, \forall r \in \left\{1,\ldots,R\right\} \quad
\textrm{and} \quad \sum_{r=1}^{R}{a_{rn}}=1.
\label{eqt:contraints_linear_model}
\end{equation}
However, in order to transform the sum to one constraint into an inequality constraint (which will be handled more easily in the algorithm), we propose the following reparametrization
\begin{equation}
a_{rn} = \left(\prod_{k=1}^{r-1} t_{kn} \right) \times \left\lbrace \begin{array}{ll}
               1-t_{rn}, & \textrm{if  } r < R \\
               1,        & \textrm{if  } r = R
             \end{array}
 \right..
\label{eqt:Abundance_tranformation}
\end{equation}
The transformation \eqref{eqt:Abundance_tranformation}  has been introduced in \cite{Betancourt2013} and has shown interesting properties for hyperspectral unmixing in \cite{Altmann2014b}. Its main advantage is to express the positivity and the sum to one constraints for the abundances as follows
\begin{equation}
0 < t_{rn} < 1, \forall r \in {1, \cdots, R-1}
\label{eqt:new_contraints}
\end{equation}
which will be easily handled in the sampling procedure developed in this paper (see Sections  \ref{sec:Hierarchical_Bayesian_model} and \ref{sec:Hybrid_Gibbs_algorithm}).

%%%%%%%%%%%%%%%%%%%%%%%%%%%%%%%%%%%%%%%%%%%%%%%%%%%
%%%%%%%%%%%%%%%%%%%%%%%%%%%%%%%%%%%%%%%%%%%%%%%%%%%
%%%%%%%%%%%%%%%%%%%%%%%%%%%%%%%%%%%%%%%%%%%%%%%%%%%
\section{Hierarchical Bayesian model} \label{sec:Hierarchical_Bayesian_model}
This section introduces a hierarchical Bayesian model for unsupervised hyperspectral SU accounting for spectral variability. The unknown parameters of this model include the ($L \times R$) endmember mean matrix $\bsM$, the ($R \times L$) matrix  $\bSig$ gathering the endmember variances  (with $\bSig_{r,l} = \sigma^2_{rl}$), the $(R-1) \times N$ reparameterized abundance matrix $\bsT$ (whose $n$th column is $\bsT_{:n}  = \bst_n$), the ($1 \times N$) label vector $\bsz$ and the ($1 \times N$) vector $\bPsi$ containing the noise variances (with $\bPsi_{n}  = \psi_n^2$).

%$\boldsymbol{\theta} = \left\lbrace \bsT, \bsM, \bSig, \bsz, \bsC  \right\rbrace$

\subsection{Likelihood} \label{subsec:Likelihood}
Using the observation model \eqref{eqt:Normal_compositional_model1}, the Gaussian
properties of both the noise sequence $\bse_n$ and the
endmembers, and exploiting independence between the observations in different spectral bands, yield
\begin{equation}
f(\bsy_{n}|\bsT, \bsM, \bSig, \bsz, \bPsi) \propto {\left( \frac{1}{ \prod_{l=1}^{L} \bOme_{ln} }\right)
}^{\frac{1}{2}} \exp
\left\lbrace -\frac{1}{2}  \bLam_{:n}^{T}  \left[ \left(\bsy_{n}-\bsM \bsa_{n}\right) \odot \left(\bsy_{n}-\bsM \bsa_{n}\right)\right]
\right\rbrace \label{eqt:likelihood}
\end{equation}
where $\bOme = \bSig^T \left(\bsA \odot  \bsA\right) + \bsK$ is an ($L \times N$) matrix, $\bsA= \left[\bsa_1,\cdots,\bsa_N\right]$ is an ($R \times N$) abundance matrix, $\bsK =  \boldsymbol{1}_L \otimes  \bPsi$ is an  ($L \times N$) matrix whose rows are equal to $\bPsi$, $\boldsymbol{1}_L$ is an  ($L \times 1$) vector of $1$, $\bLam $ is an ($L \times N$) matrix with $\bLam_{ln}  = \frac{1}{\bOme_{ln}}$, $\odot$ denotes the Hadamard (termwise) product and $\otimes$ denotes the Kronecker product. Note that the abundance vector $\bsa_{n}(\bst_n)$ has been denoted as $\bsa_{n}$ in \eqref{eqt:likelihood} for brevity. Moreover, contrary to the LMM, Eq. \eqref{eqt:likelihood} shows that the elements\footnote{The matrix $\bOme$ gathers the noise and endmember variances.} of $\bOme$ depend jointly on the pixel abundances and on the pixel index $\# n$. This property was also satisfied by the NCM model as previously shown in \cite{Eches2010ip,Zare2013}. Note finally that the joint likelihood of the observation matrix $\bsY$ can be obtained by exploiting independence between the observed pixels
\begin{equation}
f(\bsY|\bsT, \bsM, \bSig, \bsz, \bPsi) \propto
\prod_{n=1}^{N}{
f(\bsy_{n}|\bsT, \bsM, \bSig, \bsz, \bPsi)}.
\label{eqt:joint_likelihood}
\end{equation}

\subsection{Parameter priors} \label{subsec:Parameter_priors}
This section introduces the prior distributions that we have chosen for the parameters of interest  $\bsz$, $\bsT$ (or $\bsA$), $\bsM$, and $\bSig$.

\subsubsection{Classification prior modeling} \label{subsubsec:Label_prior}
Many recent works related to hyperspectral imaging have been
considering spatial correlation between the image pixels to segment the image into homogeneous regions with similar abundances \cite{Eches2010tgrs,Nascimento2012}. In this paper, we propose to exploit this correlation by dividing the observed image into $K$ classes sharing the same abundance properties \cite{Eches2010tgrs}. Each pixel is assigned to a specific class by using a latent label variable  $z_n$ that takes its value into a finite set $\left\lbrace 1,\cdots,K\right\rbrace$. The whole set of random variables $\left\lbrace z_n\right\rbrace_{n=1,\cdots,N}$  forms a random field. The correlation between neighboring pixels is then introduced by considering a Markov random field prior for $z_n$ as follows
%to relate it to its neighbors $\bsz_{\nu(n)}$ as follows
\begin{equation}
f\left(z_{n}  | \bsz_{\backslash n}\right)  =  f\left(z_{n} | \bsz_{\nu(n) }\right)
\label{eqt:priorLabels1}
\end{equation}
where $\nu(n)$ denotes the pixel neighborhood as in \cite{Eches2010tgrs} (a four neighborhood structure will be considered in the rest of the paper), $\bsz_{\nu(n) } = \left\lbrace z_i, i \in \nu(n)\right\rbrace$ and $\bsz_{\backslash n} = \left\lbrace z_i, i \neq n\right\rbrace$. As in \cite{Bali2008,Eches2010tgrs,Altmann2014}, this paper considers a Potts-Markov model which is appropriate for hyperspectral image segmentation. The prior of $\bsz$ is then
obtained using the Hammersley-Clifford theorem
\begin{equation}
f\left(\bsz\right)  =
\frac{1}{G(\beta)} \exp\left[\beta \sum_{n=1}^{N}{\sum_{n' \in \nu(n) }  \delta \left(z_{n}-z_{n'} \right)}   \right]
\label{eqt:priorLabels2}
\end{equation}
where $\beta>0$ is the granularity coefficient, $G(\beta)$ is a normalizing (or partition) constant and $\delta(.)$ is the Dirac delta function. The parameter $\beta$ controls the degree of homogeneity of each region in the image. It is assumed known a priori in this paper. However, it could  be also included within the Bayesian model and estimated using the strategy described in \cite{Pereyra2013b}.

\subsubsection{Abundance matrix $\bsT$} \label{subsubsec:Coefficient_matrix}
In order to satisfy the constraints \eqref{eqt:contraints_linear_model}, the abundance vector should live in the following simplex $\calS$
\begin{equation}
\calS = \left\lbrace \bsa_n  \big| a_{rn}\geq 0,
\forall r  \; \textrm{and} \;
\sum_{r =1}^{R}{a_{rn} } = 1 \right\rbrace.
\label{eqt:Simplex_Alpha}
\end{equation}
Thus, a natural choice for the prior of $\bsa_n$ is a uniform distribution on  $\calS$  \cite{Dobigeon2008,Halimi2011TGRS}. However, we want to define a prior enforcing strong correlations for close pixels. Therefore, we propose to assign a Dirichlet prior to the abundances of the $k$th class  of the image with Dirichlet parameters $\bsc_{k} = \left(\bsc_{1k}, \cdots,\bsc_{Rk} \right)^T$ as follows
\begin{equation}
\bsa_n | z_n=k,\bsc_{k} \sim \textrm{Dir} (\bsc_{k}), \textrm{for } n \in \mathcal{I}_k
\label{eqt:Simplex_Alpha}
\end{equation}
where $\textrm{Dir} (.)$ denotes the Dirichlet distribution, and $n\in \mathcal{I}_k$ means that $\bsy_n$ belongs to the $k$th class (which is also equivalent to $z_n=k$). This prior allows the data to be located in several different clusters inside the simplex \cite{Nascimento2012}. Note that assigning a Dirichlet prior for $\bsa_n$ corresponds to a beta distribution prior for the coefficient $t_{rn}$ as shown in \cite{Betancourt2013,Altmann2014b}
\begin{equation}
t_{rn}| z_n= k,\bsC_{r:R,k}    \sim  \calB e \left( \sum_{i=r+1}^{R}{c_{ik}}, c_{rk}\right), \textrm{for } n \in \mathcal{I}_k
\label{eqt:priorZ1}
\end{equation}
where $\bsC = \left[\bsc_1,\cdots,\bsc_K\right]$ is an $R\times K$ matrix containing the Dirichlet parameters.
The prior associated with the vector $\bst_{n}$ is finally obtained by assuming prior independence between its elements leading to
\begin{equation}
f\left(\bst_{n} | z_n=k,\bsc_{k} \right) =
 \frac{\Gamma\left(\sum_{i=1}^{R}{ c_{ik}}\right)}
 {\prod_{i=1}^{R}\Gamma \left(c_{ik} \right)}
\bone_{\left[0,1 \right]^{R-1}} \left(\bst_{n}\right)
\prod_{r=1}^{R-1}{ t_{rn}^{  \sum_{i=r+1}^{R}{ c_{ik}-1} }   \left(1-t_{rn}\right)^{c_{rk}-1}}
\label{eqt:priorZ2}
\end{equation}
for $n \in \mathcal{I}_k$, where $\bone_{\left[0,1 \right]^{R-1}} (.)$ is the indicator of the set $\left[0,1 \right]^{R-1}$.

%\textcolor[rgb]{0.98,0.00,0.00}{To check or change if error, product of beta et non beta seule}

%is equivalent to assigning a Beta prior for the coefficient vector $\bst_n$ as follows
%
%Moreover, we will assume in the present paper that
%
%Dirichlet distribution with different coefficients for each spatial class
%\begin{equation}
%t_{rn}| z_n= k,\bsC_{r:R,k}    \sim  \calB e \left( \sum_{i=r+1}^{R}{c_{ik}}, c_{rk}\right), \textrm{for } n \in \mathcal{I}_k
%\label{eqt:priorZ1}
%\end{equation}

\subsubsection{Endmember means} \label{subsubsec:Mean_of_endmembers}
The endmember mean matrix $\bsM$ contains reflectances that should satisfy the following constraints \cite{Altmann2014b}
\begin{equation}
0< \bsm_{rl} <1, \forall r \in \left\lbrace1,\cdots,R\right\rbrace, \forall l \in\left\lbrace1,\cdots,L\right\rbrace.
\label{eqt:constraints_M}
\end{equation}
Moreover, it makes sense to assume that the reflectances are close to estimates computed using an EEA. Therefore, we choose a truncated Gaussian prior for each endmember as follows \cite{Zare2013,Altmann2014b}
\begin{equation}
\bsm_{r} \sim \calN_{[0,1]^L} \left(\widetilde{\bsm}_{r}, \epsilon^2 \mathds{I}_l \right)
\label{eqt:priorM}
\end{equation}
where $\widetilde{\bsm}_{r}$ denotes an estimated endmember (resulting from an EEA such as VCA\footnote{We consider in this paper the VCA algorithm even if other algorithms such as N-FINDR \cite{Winter1999} and \textit{pixel purity index} (PPI) \cite{Boardman1993}  could also be investigated.}) and  $\epsilon^2$ is a variance term defining the confidence that we have on this estimated endmember $\widetilde{\bsm}_{r}$.

\subsubsection{Endmember variances} \label{subsubsec:Variance_of_endmembers}
The absence of knowledge about the endmember variances can be considered by choosing a Jeffreys distribution for the parameters $\sigma^2_{rl}$, i.e.,
\begin{equation}
f\left(\bSig_{:l}\right) \propto  \prod_{r=1}^{R} \frac{1}{\sigma^2_{rl}}   \bone_{\dsR+} \left(\sigma^2_{rl}\right)  \label{eqt:priorSigma2}
\end{equation}
where we have assumed prior independence between the endmember variances.

\subsubsection{Noise variance prior} \label{subsubsec:Label_prior}
To avoid identifiability problems, the noise effect should be smaller than the effect of endmember variability. This can be achieved by choosing an exponential prior
\begin{equation}
f\left(\psi_{n}^2  | \lambda\right)  =  \lambda  \exp{\left(-\lambda \psi_{n}^2 \right)} \bone_{\dsR+} \left(\psi_{n}^2\right)
\label{eqt:priorNoise}
\end{equation}
where $\lambda$ is a large coefficient imposing sparsity for $\psi_{n}$ ($\lambda= 10^{7}$ in our simulations). We furthermore assume prior independence between the random variables $\psi_{n}^2, \forall n \in \left\lbrace1,\cdots,N \right\rbrace$.
%This prior ensures small values for $\psi_{n}^2$ since the  randomness of an hyperspectral image is assumed to be mainly due to endmember variability.
Note that the estimation of $\psi_{n}^2$ can be removed from the proposed Bayesian algorithm without changing significantly the estimation  performance (see Section \ref{subsec:Comparison_with_other_algorithms}). This paper presents a  general formulation allowing the noise effect to be removed by setting to zero the noise variance.

\subsection{Hyperparameter priors} \label{subsec:Hyperparameter_priors}

\subsubsection{Dirichlet parameters} \label{subsubsec:Dirichlet_coefficients}
The Dirichlet parameters $\bsc_k$ are assigned the following conjugate prior \cite{Zhanyu_Eusipco2012}
\begin{equation}
f\left(\bsc_{k}| z_n = k\right)  =   \left[\frac{\Gamma\left(\sum_{r=1}^{R}{ c_{rk}}\right)}
 {\prod_{r=1}^{R}\Gamma \left(c_{rk} \right)} \right]^{\gamma} \exp{\left(-\alpha \sum_{r=1}^{R}{c_{rk}} +R \alpha \right)}  \prod_{r=1}^{R}{\bone_{\dsR+} \left(c_{rk}\right)}
\label{eqt:priorC}
\end{equation}
where $\alpha$ and $\gamma$ are fixed constants that have been chosen to ensure a non-informative prior (flat distribution).

\subsection{Posterior distribution} \label{subsec:Posterior_distribution}
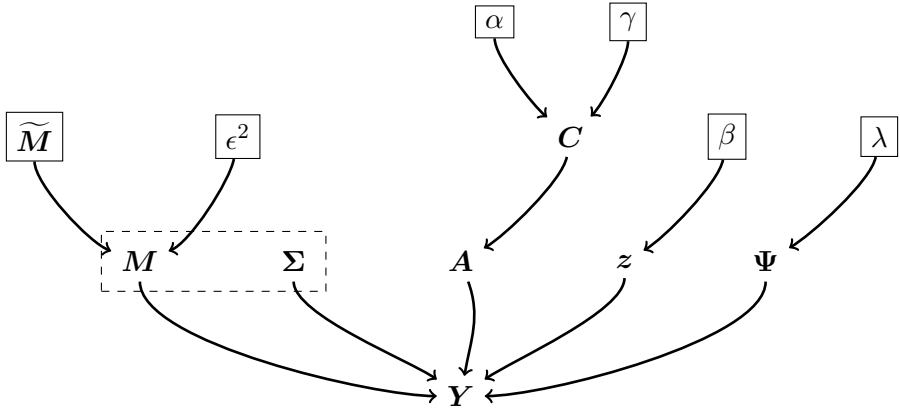
\begin{figure}
\centering
\begin{tikzpicture}
 nodes %
\node[text centered] (Y) {$\bsY$};
\node[above =1.25 of Y, text centered] (A) {$\bsA$};
\node[left  =1.6  of A, text centered] (Sig) {$\bSig$};
\node[right =1.6  of A, text centered] (h) {$\bsz$};
\node[right =3.4  of A, text centered] (Psi) {$\bPsi$};
\node[above =1.25  of A, text centered] (t) {$$};
\node[right =1.  of t, text centered] (C) {$\bsC$};
\node[draw, rectangle, right =1.5  of C, text centered] (B) {$\beta$};
\node[draw, rectangle,right =1.5  of B, text centered] (lam) {$\lambda$};
%\node[left  =0.9    of t, text centered] (Sig) {$\bSig$};
\node[left  =1.35  of Sig, text centered] (M) {$\bsM$};
\node[above =1.25  of t, text centered] (tt) {$$};
\node[draw, rectangle, right =0.05 of tt, text centered] (x1) {$\alpha$};
\node[draw, rectangle, right =1.25  of x1, text centered] (x2) {$\gamma$};
\node[draw, rectangle, left  =2.5  of t, text centered] (ep) {$\epsilon^2$};
\node[draw, rectangle, left  =2.  of ep, text centered] (Mt) {$\widetilde{\bsM}$};
\node[draw,dashed,fit=(M) (Sig)] {};

 edges %
\draw[->, line width= 1] (A) to  [out=290,in=80, looseness=1] (Y);
\draw[->, line width= 1] (Sig) to  [out=270,in=145, looseness=0.5] (Y);
\draw[->, line width= 1] (h) to  [out=270,in=35, looseness=0.5] (Y);
\draw[->, line width= 1] (Psi) to  [out=270,in=0, looseness=0.5] (Y);
\draw[->, line width= 1] (C) to  [out=260,in=30, looseness=0.5] (A);
\draw[->, line width= 1] (lam) to  [out=260,in=30, looseness=0.5] (Psi);
\draw[->, line width= 1] (B) to  [out=260,in=30, looseness=0.5] (h);
%\draw[->, line width= 1] (Sig) to  [out=270,in=45, looseness=0.5] (S);
\draw[->, line width= 1] (M) to  [out=270,in=180, looseness=0.5] (Y);
\draw[->, line width= 1] (x1) to  [out=270,in=140, looseness=0.5] (C);
\draw[->, line width= 1] (x2) to  [out=270,in=45, looseness=0.5] (C);
\draw[->, line width= 1] (ep) to  [out=260,in=20, looseness=0.5] (M);
\draw[->, line width= 1] (Mt) to  [out=270,in=160, looseness=0.5] (M);
\end{tikzpicture}
\caption{DAG for the parameter and hyperparameter priors (the fixed parameters appear in boxes). Note that the dashed box defines the statistical distribution of the endmember matrix $\bsS$.}
\label{fig:DAG}
\end{figure}

The parameters of the proposed Bayesian model are included in the vector  $\bthe = \left\lbrace \bthe_p,\bthe_h\right\rbrace$ where $\bthe_p = \left\lbrace\bsT, \bsM, \bSig, \bsz, \bPsi\right\rbrace$ (parameters) and  $\bthe_h = \left\lbrace\bsC\right\rbrace$ (hyperparameters). This Bayesian model is summarized in the directed acyclic graph (DAG) displayed in Fig. \ref{fig:DAG}.

The joint posterior distribution of the unknown parameter/hyperparameter vector $\bthe$ can be computed from the following hierarchical structure
\begin{equation}
f\left(\bthe_p,\bthe_h | \bsY \right)  \propto f\left(\bsY | \bthe_p,\bthe_h  \right) f\left(\bthe_p,\bthe_h  \right)
\label{eqt:Bayes}
\end{equation}
where $f\left(\bsY | \bthe_p,\bthe_h  \right) =f\left(\bsY | \bthe_p\right) $ has been defined in \eqref{eqt:joint_likelihood} and $f\left(\bthe_p,\bthe_h  \right)$ is the joint prior of the unknown parameters. Assuming prior independence between the parameters yields
\begin{eqnarray}
f\left(\bthe_p,\bthe_h  \right) & = & f\left( \bthe_p | \bthe_h  \right) f\left(\bthe_h  \right)  \nonumber \\
& = & f\left( \bsT| \bsC  \right) f\left( \bsM \right) f\left( \bSig \right) f\left( \bsz \right) f\left( \bPsi \right) f\left( \bsC  \right).
\label{eqt:Prior}
\end{eqnarray}
The joint posterior distribution $f\left(\bthe_p,\bthe_h | \bsY \right)$ can be computed up to a multiplicative
constant after replacing \eqref{eqt:joint_likelihood} and \eqref{eqt:Prior} in \eqref{eqt:Bayes}. Unfortunately, it is difficult to obtain closed form expressions for the standard Bayesian estimators associated with \eqref{eqt:Bayes}. In this paper, we propose to use MCMC methods to generate samples asymptotically distributed according to \eqref{eqt:Bayes} and to build estimators of $\bthe$ from these generated samples. Due to the large number of parameters to be sampled, we use an HMC algorithm which improves the mixing properties of the sampler  and  reduces the required number of iterations to approximate the target distribution \cite{Brooks2011}. The parameters are finally estimated using the minimum mean square error (MMSE) estimator for $\left\lbrace\bsT, \bsM, \bSig, \bPsi, \bsC\right\rbrace$   and the maximum a posteriori (MAP) estimator for the labels $\bsz$. The next section defines the proposed sampling procedure based on a hybrid Gibbs sampler including a CHMC method.

%%%%%%%%%%%%%%%%%%%%%%%%%%%%%%%%%%%%%%%%%%%%%%%%%%%
%%%%%%%%%%%%%%%%%%%%%%%%%%%%%%%%%%%%%%%%%%%%%%%%%%%
%%%%%%%%%%%%%%%%%%%%%%%%%%%%%%%%%%%%%%%%%%%%%%%%%%%
\section{Hybrid Gibbs algorithm} \label{sec:Hybrid_Gibbs_algorithm}
The principle of the Gibbs sampler is to generate samples according to the conditional distributions of the target distribution (here the posterior \eqref{eqt:Bayes}) \cite{Robert2007}. When a conditional distribution cannot be sampled directly, sampling techniques such as the Metropolis-Hasting (MH) algorithm can be applied. In this paper, we consider HMC as the proposal strategy since it provides better mixing property than independent or random walk MH moves especially for high-dimensional problems. The next section describes the CHMC algorithm followed by the description of the sampling procedure of the conditional distributions.

\subsection{Constrained Hamiltonian Monte Carlo method} \label{subsec:Constrained_Hamiltonian_Monte_Carlo_method}
%-Introduire un moment
%-calcul de Hamiltonien
%-dire qu'on obtient des EDP qui sont résolues par les 3 eqts
%-introduction des contraintes par symétrisation
%-voir algo.

HMC is used to sample the high dimensional parameter vector of the proposed Bayesian model. It exploits the gradient of the target distribution to improve the quality of the generated samples. Denoting as $f(\bsq)$ (resp. $\bsq$) the distribution (resp. d-dimensional variable) to be sampled from, HMC defines the Hamiltonian function after introducing a Gaussian momentum variable $\bsp$ as follows
\begin{equation}
H \left(\bsp,\bsq\right) = U(\bsq) + K(\bsp)
\label{eqt:Hamiton}
\end{equation}
where $U(\bsq) = -\log\left[f(\bsq)\right]$ is the potential energy related to the target distribution $f(\bsq)$ and $K(\bsp)  = \frac{1}{2}\bsp^T\bsp $ is the momentum energy which results from an independent centered Gaussian distribution for $\bsp$ \cite{Altmann2014b}.
The evolution of the $(\bsq, \bsp)$ samples is determined using the partial derivatives of the Hamiltonian referred to as Hamiltonian equations \cite{Brooks2011}. For computer implementations, these equations should be discretized which can be done using the leapfrog method that ensures volume preservation and reversibility of the chains. This leapfrog  discretization scheme moves the samples by an $\epsilon$ stepsize, i.e., from the $n$th state $\left(\bsq^{n},\bsp^{n}\right)$  to the $(n+1)$th state $\left(\bsq^{(n+1)},\bsp^{(n+1)}\right)$ using $N_{L}$ iteration steps  defined by
\begin{align}
\bsp^{(i,n+1/2)} & =  \bsp^{(i,n)} - \frac{\epsilon}{2}  \frac{\partial U}{\partial \bsq^T} \left[ \bsq^{(i,n)} \right]   \\
\bsq^{(i,n+1)} & =  \bsq^{(i,n)} + \epsilon  \bsp^{(i,n+1/2)}  \\
\bsp^{(i,n+1)} & =  \bsp^{(i,n+1/2)} - \frac{\epsilon}{2}  \frac{\partial U}{\partial \bsq^T} \left[ \bsq^{(i,n+1)} \right].
\label{eqt:LF_steps}
\end{align}
The resulting samples are accepted with probability $\rho$ given by
\begin{equation}
\rho = \min\left\lbrace 1, \exp\left[H\left(\bsq^{n},\bsp^{n}\right)  - H\left(\bsq^{(n+1)},\bsp^{(n+1)}\right)  \right] \right\rbrace.
\label{eqt:proba_acceptance}
\end{equation}
This procedure ensures the resulting samples to be asymptotically distributed according to the target distribution.

In the presence of inequality constraints ($\bsq^{(i,n\epsilon)} \in \left[q_l,q_u\right]$), we adopt the procedure presented in \cite{Altmann2014b} and \cite[Chap.~5]{Brooks2011}. This procedure replaces a sample that violates the constraints at each leapfrog iteration by its symmetric to the bound (see \cite{Altmann2014b} for more details). For example, the candidate  $\bsq^{(i,n)} = q_u + h$ with $0< h< (q_u-q_l)$  will be replaced by $\bsq^{(i,n)} = q_u - h$ (and similarly  $\bsq^{(i,n)} = q_l - h$ will be replaced by $\bsq^{(i,n)} = q_l + h$) when a constraint is not  satisfied.

\subsection{Sampling the abundance matrix $\bsT$} \label{subsec:Sampling_the_coefficient_matrix_Z}
It can be shown that the $N$ vectors $\bst_{n}, n \in \left\lbrace1,\cdots,N \right\rbrace$ are a posteriori independent leading to
\begin{equation}
f\left(\bsT| \bsY, \bsM, \bSig, \bsC\right) = \prod_{k=1}^{K}{ \prod_{n\in \mathcal{I}_k }{f\left(\bst_{n}| z_n=k, \bsy_{n}, \bsM, \bSig, \bsc_{k}\right)} }. \label{eqt:posteriorZindep}
\end{equation}
Moreover, using the likelihood \eqref{eqt:likelihood} and the prior \eqref{eqt:priorZ2} leads to the following conditional distribution
\begin{eqnarray}
f\left(\bst_{n}| z_n=k, \bsy_{n}, \bsM, \bSig, \bsc_{k}\right) & \propto & {\left( \frac{1}{ \prod_{l=1}^{L} \bOme_{ln} }\right)
}^{\frac{1}{2}} \exp
\left\lbrace -\frac{1}{2}  \bLam_{:n}^{T}  \left[ \left(\bsy_{n}-\bsM \bsa_{n}\right) \odot \left(\bsy_{n}-\bsM \bsa_{n}\right)\right]
\right\rbrace  \nonumber \\
& \times &  \bone_{\left[0,1 \right]^{R-1}} \left(\bst_{n}\right) \prod_{r=1}^{R-1}{ t_{rn}^{  \sum_{i=r+1}^{R}{ c_{ik}-1} }   \left(1-t_{rn}\right)^{c_{rk}-1}
} \label{eqt:posteriorZ}
\end{eqnarray}
for $n \in \mathcal{I}_k$. The conditional distribution \eqref{eqt:posteriorZ} is not easy to sample. However, the CHMC framework is well suited for sampling the independent vectors $\bst_{n}, n \in \left\lbrace1,\cdots,N \right\rbrace$ in an effective parallel procedure that reduces the computational cost. Moreover, the small size of these vectors (of size $(R-1) \times 1$) improves the convergence of the sampler. Note that the CHMC requires the definition of the potential energy $U\left(\bst_{n}\right) = -\log\left[f\left(\bst_{n}| z_n=k, \bsy_{n}, \bsM, \bSig, \bsc_{k}\right) \right] $ given by
\begin{equation}
U \left(\bst_{n}\right) = U_1 + U_2 + U_3
\label{eqt:HamitonZ1}
\end{equation}
with
\begin{eqnarray}
U_1 & = & \frac{1}{2}  \bLam_{:n}^{T}  \left[ \left(\bsy_{n}-\bsM \bsa_{n}\right) \odot \left(\bsy_{n}-\bsM \bsa_{n}\right)\right] \nonumber \\
U_2 & = & -\sum_{r=1}^{R}{ \left\lbrace \left(\sum_{i=r+1}^{R}{ c_{ik}-1} \right)  \log \left(t_{rn}\right) +  \left(c_{rk}-1 \right) \log\left(1-  t_{rn}  \right)   \right\rbrace }
\nonumber \\
U_3 & = & \frac{1}{2} \sum_{l=1}^{L}{ \log \left(\bOme_{ln}\right) }.
\label{eqt:HamitonZ2}
\end{eqnarray}
Note finally that the derivatives of $U$ with respect to the variable of interest $\bst_n$  (that are required for the CHMC steps)  are provided in the appendix.

\subsection{Sampling the mean endmember matrix $\bsM$} \label{subsec:Sampling_the_endmember_matrix_M}
Straightforward computations using the posterior distribution \eqref{eqt:Bayes}  yield
\begin{equation}
f\left(\bsM| \bsY, \bsT, \bSig\right) = \prod_{l=1}^{L}{f\left(\bsM_{l:}| \bsY_{l:}, \bsT, \bSig_{:l}\right)} \label{eqt:posteriorM_indep}
\end{equation}
where
\begin{eqnarray}
f\left(\bsM_{l:}| \bsY_{l:}, \bsT, \bSig_{:l}\right) & \propto &   \exp
\left\lbrace -\frac{1}{2}   \left[ \left(\bsY_{l:}-\bsM_{l:} \bsA\right) \odot \left(\bsY_{l:}-\bsM_{l:} \bsA\right)\right] \bLam_{l:}^{T}
\right\rbrace \nonumber \\
& \times &
\exp \left( -\frac{|| \bsM_{l:} - \widetilde{\bsM}_{l:} ||^2}{2 \epsilon^2} \right) \bone_{\left[0,1 \right]^R} \left(\bsM_{l:}\right).  \label{eqt:posteriorM}
\end{eqnarray}
Equation \eqref{eqt:posteriorM_indep} results from the independence between the columns of the matrix $\bsM$ (vectors of small size $R\times 1$). This interesting property promotes the use of a parallel CHMC algorithm for sampling $\bsT$. The potential energy $V$ associated with the conditional distribution of $\bsM_{l:}$ is given by
\begin{equation}
V \left(\bsM_{l:}\right) =  \frac{1}{2}   \left[ \left(\bsY_{l:}-\bsM_{l:} \bsA\right) \odot \left(\bsY_{l:}-\bsM_{l:} \bsA\right)\right] \bLam_{l:}^{T}
+ \frac{|| \bsM_{l:} - \widetilde{\bsM}_{l:} ||^2}{2 \epsilon^2}.
\label{eqt:HamitonM}
\end{equation}
The derivatives of $V$ with respect to $\bsM_{l:}$ are provided in the appendix.

\subsection{Sampling the variance of the endmember matrix} \label{subsec:Sampling_the_variance_of_the_endmember_matrix}
Considering \eqref{eqt:Bayes} yields the following conditional distribution for matrix $\bSig$ containing the endmember variances
\begin{equation}
f\left(\bSig | \bsY, \bsT, \bsM\right) =  \prod_{l=1}^{L}{f\left(\bSig_{:l} | \bsY_{l:}, \bsT, \bsM_{l:}\right)} \label{eqt:posteriorSigma}
\end{equation}
with
\begin{eqnarray}
f\left(\bSig_{:l} | \bsY_{l:}, \bsT, \bsM_{l:}\right) & \propto & {\left( \frac{1}{ \prod_{n=1}^{N} \bOme_{ln} }\right)}^{\frac{1}{2}}
\exp \left\lbrace -\frac{1}{2}   \left[ \left(\bsY_{l:}-\bsM_{l:} \bsA\right) \odot \left(\bsY_{l:}-\bsM_{l:} \bsA\right)\right] \bLam_{l:}^{T}
\right\rbrace \nonumber \\
& \times & \prod_{r=1}^{R} {\frac{1}{\sigma^2_{rl}}  \bone_{\dsR+} \left(\sigma^2_{rl}\right)}. \label{eqt:posteriorSigma2}
\end{eqnarray}
Sampling from \eqref{eqt:posteriorSigma2} can again be performed using a CHMC algorithm (as in Sections \ref{subsec:Sampling_the_coefficient_matrix_Z} and \ref{subsec:Sampling_the_endmember_matrix_M}). The potential energy associated with the vector $\bSig_{:l}$ is
\begin{equation}
W \left(\bSig_{:l}\right) = W_1 + W_2 + W_3
\label{eqt:HamitonSig}
\end{equation}
with
\begin{eqnarray}
W_1 & = & \frac{1}{2}   \left[ \left(\bsY_{l:}-\bsM_{l:} \bsA\right) \odot \left(\bsY_{l:}-\bsM_{l:} \bsA\right)\right] \bLam_{l:}^{T}
 \nonumber \\
W_2 & = & \sum_{r=1}^{R}{\log \left(\sigma^2_{rl}\right) }
\nonumber \\
W_3 & = &  \frac{1}{2} \sum_{n=1}^{N}{ \log \left(\bOme_{ln}\right) }.
\label{eqt:HamitonZ2}
\end{eqnarray}
The derivatives of $W$ with respect to $\bSig_{:l}$ are provided in the appendix.

\subsection{Sampling the labels} \label{subsec:Sampling_the_labels}

The conditional distribution associated with the discrete random variable $z_n$ is given by
\begin{equation}
f\left(z_n = k| \bst_{n}, \bsc_{k} \right)  \propto
f\left(\bst_{n} | z_n=k, \bsc_{k} \right) \,
\exp\left[\beta \sum_{n' \in \nu(n) }  \delta \left(k - z_{n'} \right)   \right]
\label{eqt:posteriorlabels}
\end{equation}
where $f\left(\bst_{n} | z_n=k, \bsc_{k}\right)$ has been defined in \eqref{eqt:priorZ2}. Sampling from this conditional distribution is
classically performed by drawing a discrete value in the finite set $\left\lbrace 1,\cdots,K \right\rbrace$ with the probabilities \eqref{eqt:posteriorlabels}.

\subsection{Sampling the noise variance $\bPsi$}
Considering \eqref{eqt:Bayes} yields the following conditional distribution for the noise variance matrix $\bPsi$
\begin{equation}
f\left(\bPsi| \bsz, \bsT, \bsY, \bsM, \bSig, \bsc\right) =  \prod_{n=1}^{N}{f\left(\psi_{n}^2| z_n=k, \bst_{n}, \bsy_{n}, \bsM, \bSig, \bsc_{k}\right)} \label{eqt:posteriorSigma}
\end{equation}
with
\begin{eqnarray}
f\left(\psi_{n}^2| z_n=k, \bst_{n}, \bsy_{n}, \bsM, \bSig_{:l}, \bsc_{k}\right) & \propto & {\left( \frac{1}{ \prod_{l=1}^{L} \bOme_{ln} }\right)
}^{\frac{1}{2}} \exp
\left\lbrace -\frac{1}{2}  \bLam_{:n}^{T}  \left[ \left(\bsy_{n}-\bsM \bsa_{n}\right) \odot \left(\bsy_{n}-\bsM \bsa_{n}\right)\right]
\right\rbrace  \nonumber \\
& \times &
\exp{\left(-\lambda \psi_{n}^2 \right)} \; \bone_{\dsR+} \left(\psi_{n}^2\right) \label{eqt:posteriorpsi}
\end{eqnarray}
for $n \in \mathcal{I}_k$. This distribution is sampled using a parallel CHMC procedure with the following potential energy
\begin{equation}
H\left(\psi_{n}^2\right) = U_1 + U_3 + \lambda \psi_{n}^2.
\label{eqt:Hamiton_eps}
\end{equation}

\subsection{Sampling the Dirichlet coefficients} \label{subsec:Sampling_the_Dirichlet_coefficients}
Using \eqref{eqt:Bayes} and \eqref{eqt:Prior}, it can be easily shown that the conditional distribution of $\bsc_{k}| \bsT, \bsz_{n \in \mathcal{I}_k}$ is given by
\begin{equation}
f\left(\bsc_{k}| \bsT, \bsz_{n \in \mathcal{I}_k} \right)  \propto  \prod_{n \in  \mathcal{I}_k}{ \left\lbrace \left[\frac{\Gamma\left(\sum_{r=1}^{R}{ c_{rk}}\right)}
 {\prod_{r=1}^{R}\Gamma \left(c_{rk} \right)} \right]^{\gamma+1} \exp{\left(-\alpha \sum_{r=1}^{R}{c_{rk}} +R \alpha \right)}
\prod_{r=1}^{R}{a_{rn}^{c_{rk}-1}} \right\rbrace}
\label{eqt:posteriorC}
\end{equation}
for $k \in \left\lbrace 1, \cdots, K\right\rbrace$. This distribution is also sampled using a CHMC procedure. The corresponding potential energy is given by
\begin{equation}
P\left(\bsc_{k}\right) =  P_1 + P_2
\label{eqt:Hamiton_Dir}
\end{equation}
with
\begin{eqnarray}
P_1 & = & \left(\gamma+1 \right) \sum_{n \in  \mathcal{I}_k}{  \left[
-  \textrm{log}\Gamma\left(\sum_{r=1}^{R}{ c_{rk}}\right) + \sum_{r=1}^{R}{\textrm{log}\Gamma \left(c_{rk} \right)}
\right]  }
 \nonumber \\
P_2 & = &     \sum_{n \in  \mathcal{I}_k}{\left[
\alpha \sum_{r=1}^{R}{c_{rk}} -R \alpha  - \sum_{r=1}^{R}{\log\left(a_{rn}^{c_{rk}-1}\right)}
 \right]}.
\label{eqt:HamitonZ2}
\end{eqnarray}
%where $\textrm{ln}\Gamma (x) = \log\left[\Gamma (x)\right] $ denotes the logarithm of the gamma function.

%\textcolor[rgb]{0.98,0.00,0.00}{To check or change if error}

%%%%%%%%%%%%%%%%%%%%%%%%%%%%%%%%%%%%%%%%%%%%%%%%%%%
%%%%%%%%%%%%%%%%%%%%%%%%%%%%%%%%%%%%%%%%%%%%%%%%%%%
%%%%%%%%%%%%%%%%%%%%%%%%%%%%%%%%%%%%%%%%%%%%%%%%%%%
%\clearpage
\section{Simulation results on synthetic data} \label{sec:Simulation_results_on_synthetic_data}
This section evaluates the performance of the proposed algorithm with synthetic data. It is divided into four parts whose objectives are: 1)
introduce the criteria used for the evaluation of the unmixing quality, 2) present the different parameters that are estimated in the proposed unmixing approach, 3) analyze the behavior of the proposed algorithm as a function of the number of endmembers and the size of the image, 4) compare the proposed strategy with other state-of-the-art algorithms from the literature.

\subsection{Evaluation criteria} \label{subsec:Evaluation_criteria}
Abundances and endmembers are known for synthetic images. In this case, the quality of the unmixing strategy can be measured by comparing the estimated and actual abundances
by using the average root mean square error (aRMSE) defined by
\begin{equation}
\textrm{aRMSE}\left(\bsA\right) = \sqrt{\frac{1}{N\, R}\sum_{n=1}^{N}
\left\| \bsa_n-\hat{\bsa}_n \right\|^{2}}
\label{eqt:RMSE}
\end{equation}
where $ ||\cdot|| $ denotes the standard $l_2 $
norm such that $ ||\bsx||^2 = \bsx^T \bsx$.
The mean of the $r$th estimated endmember can be compared with the actual one by using RMSE$\left(\bsm_r\right)$ or the spectral angle mapper $\textrm{SAM}\left(\bsm_r\right)$ defined as follows
\begin{equation}
\textrm{RMSE} \left(\bsm_r\right)  = \frac{1}{\sqrt{L}}   \left\| \hat{\bsm_r}-\bsm_r  \right\|, \textrm{    } \textrm{SAM} \left(\bsm_r\right)= \arccos \left(
\frac{\hat{\bsm}_r^T \bsm_r}{\left\|\bsm_r\right\| \; \left\|\hat{\bsm}_r  \right\|}\right)
 \end{equation}
where $\arccos(\cdot)$ is the inverse cosine operator.
Moreover, the global endmember error is evaluated by the averaged RMSE (aRMSE) and averaged SAM (aSAM) given by
\begin{equation}
\textrm{aRMSE} \left(\bsM\right)=  \sqrt{\frac{1}{R}  \sum_{r=1}^{R} \left[\textrm{RMSE} \left(\bsm_r\right)\right]^2}, \textrm{    } \textrm{aSAM}\left(\bsM\right) = \frac{1}{R}  \sum_{r=1}^{R} \textrm{SAM}\left(\bsm_r\right).
\end{equation}
Note finally that the RE and SAM criteria can also be evaluated for the $\#p$th measured and estimated pixel spectra $\bsy_n$, $\hat{\bsy}_n$ as follows
\begin{equation}
\textrm{RE} =  \sqrt{\frac{1}{N\,L}  \sum_{n=1}^{N} \left\| \hat{\bsy}_n-\bsy_n  \right\|^2}, \textrm{    } \textrm{SAM} = \frac{1}{N}  \sum_{n=1}^{N} \arccos \left(
\frac{\hat{\bsy}_n^T \bsy_n}{\left\|\bsy_n\right\| \; \left\|\hat{\bsy}_n  \right\|}\right).
\end{equation}

%RMSE
%RE$_i$
%SAM$_i$
%ARE
%ASAM
%RE
%SAM

\subsection{Performance of the proposed algorithm} \label{subsec:Performance_of_the_proposed_algorithm}
\begin{figure}[h!]
\centering
\includegraphics[width=0.75\figwidth]{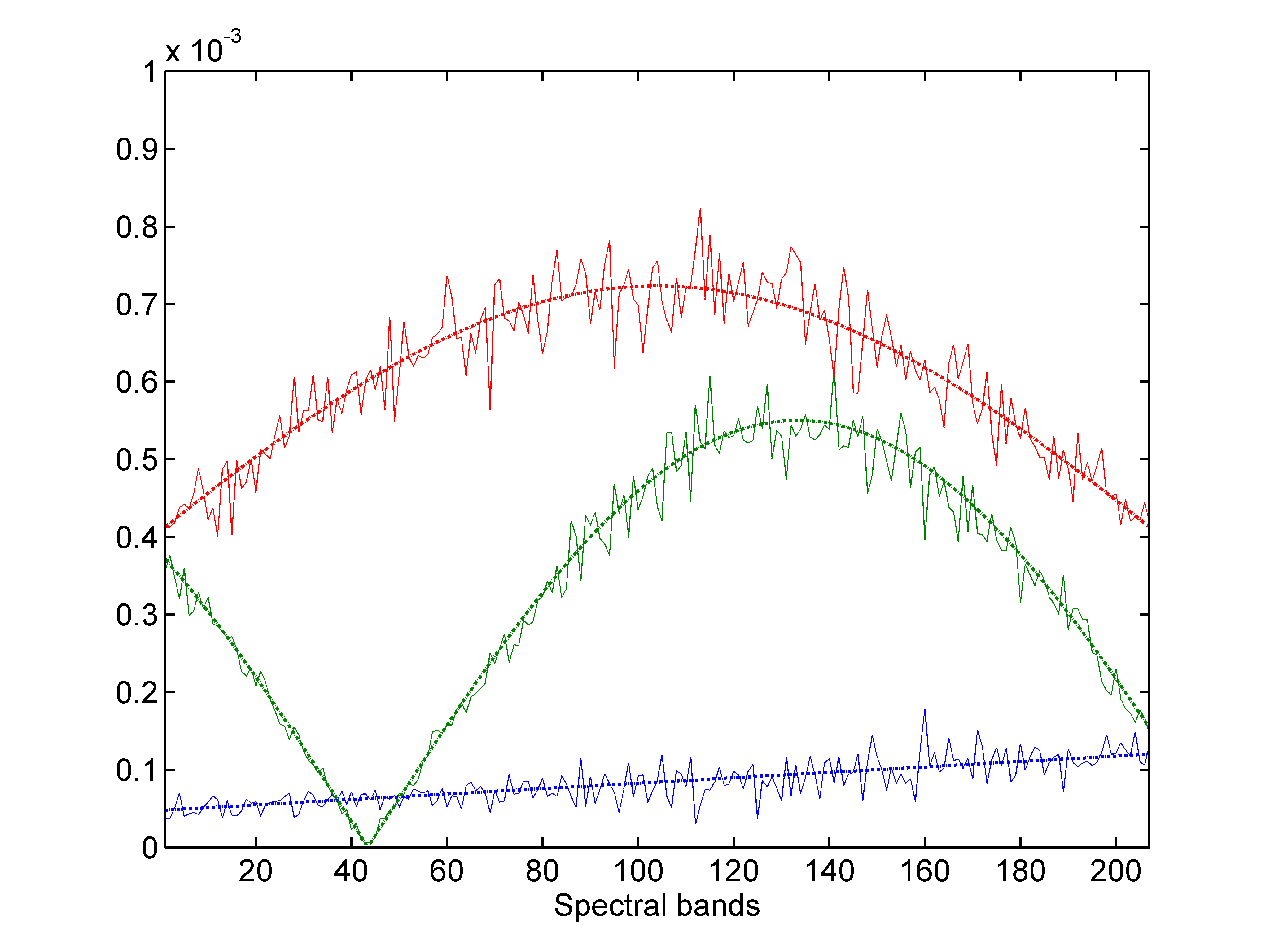}
\caption{Actual endmember variances (dassed line) and estimated variances by the proposed UsGNCM (continuous line) for the considered $R=3$ endmembers.} \label{fig:Variation_variance_Endmembers_NCM_SigmaLR_K3}
\end{figure}

\begin{figure}[h!]
\centering
\includegraphics[width=0.75\figwidth]{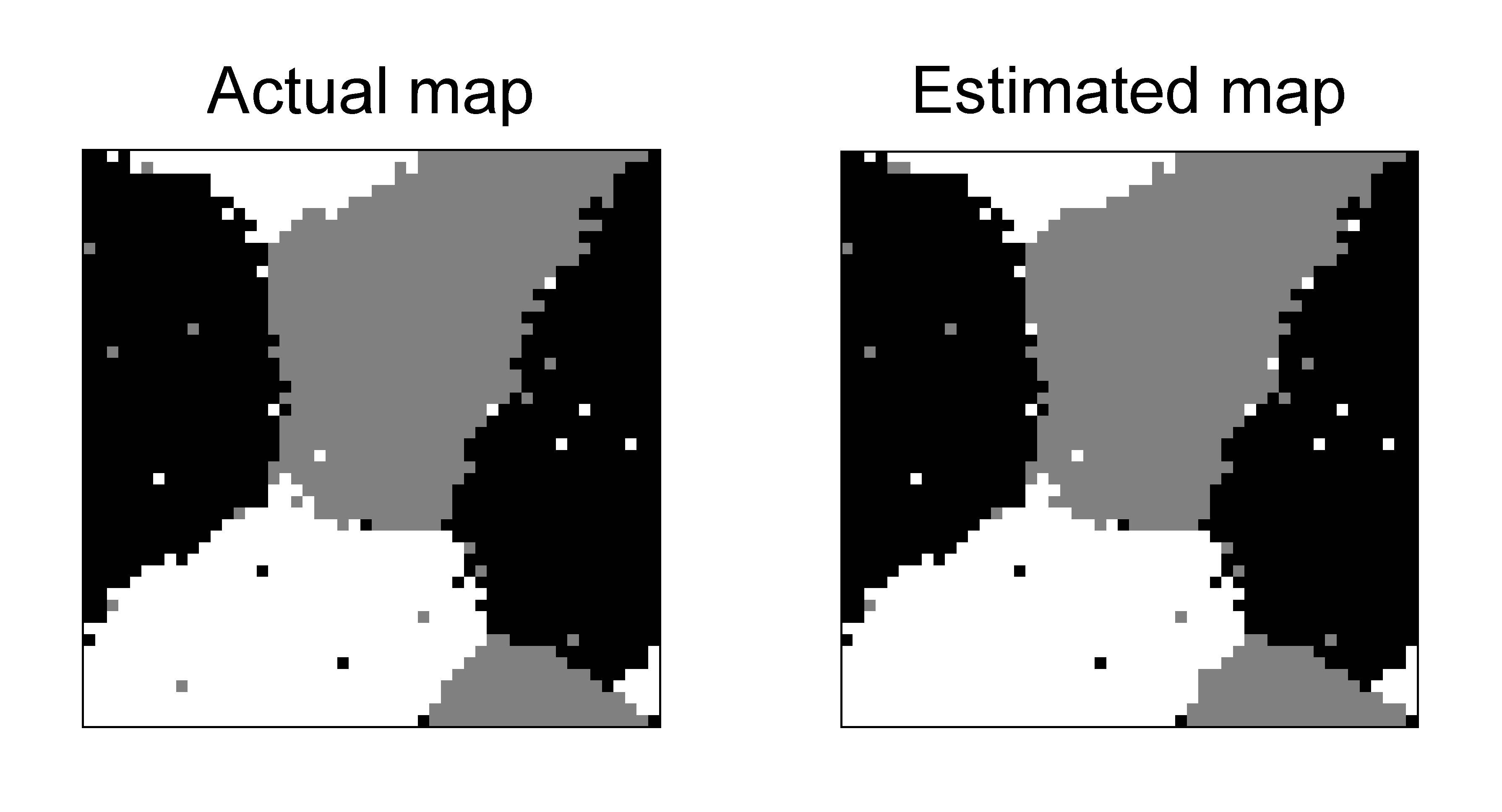}
\caption{Actual (left) and estimated (right) classification maps of a synthetic image.} \label{fig:labels_NCM_SigmaLR_epsN_K3_UsNCM}
\end{figure}

\begin{figure}[h!]
\centering
\includegraphics[width=0.75\figwidth]{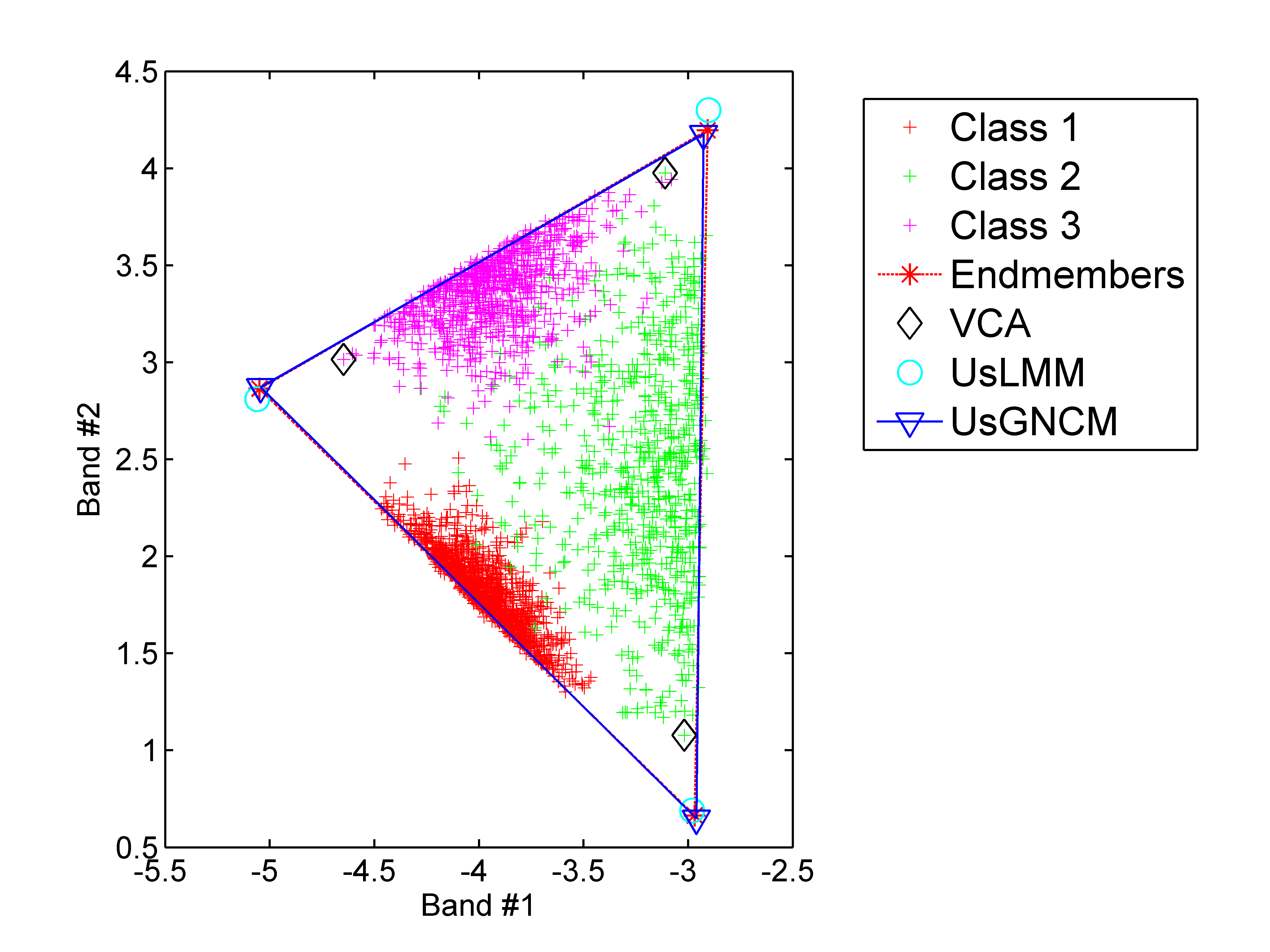}
\caption{Classified projected pixels (colored crosses), actual endmembers (red stars), endmembers estimated by VCA (black diamonds), endmembers estimated by UsLMM (cyan circle) and endmembers estimated by UsGNCM (blue triangles).} \label{fig:Projection_Data_VCA_UsNCM_epsN_K3}
\end{figure}

This section considers a $50 \times 50$ synthetic image generated according to  \eqref{eqt:Normal_compositional_model1} with $R=3$  endmembers (construction concrete, green grass and micaceous loam) that have been extracted from the ENVI software library \cite{ENVImanual2003}. The considered endmember variances depend on the spectral bands as shown in  Fig. \ref{fig:Variation_variance_Endmembers_NCM_SigmaLR_K3} (dashed lines).
This image contains $K=3$ classes whose label maps have been generated using \eqref{eqt:priorLabels2} with $\beta=1.5$ (see Fig. \ref{fig:labels_NCM_SigmaLR_epsN_K3_UsNCM}).
The abundances of each class share the same Dirichlet parameters (that are reported in Table \ref{tab:Dirichlet_coeff}) leading to the observed pixels displayed in Fig. \ref{fig:Projection_Data_VCA_UsNCM_epsN_K3}. Note that the generated abundances have been truncated ($a_r<0.9, \forall r$) to avoid the presence of pure pixels in the image. Finally, we have considered a noise variance equal to $10^{-7}$ (note that the noise variance has to be smaller than the endmember variances). The proposed unsupervised GNCM-based algorithm, denoted by UsGNCM, has been
run using $N_{\textrm{bi}} = 11000$ burn-in iterations and $N_{\textrm{MC}} = 12000$ iterations.
\renewcommand{\arraystretch}{1.1}
\begin{table}[h] \centering
\centering \caption{Actual and estimated Dirichlet parameters in each spatial class.}
\begin{tabular}{|c|c|c|c|c|c|c|}
  \cline{2-7}
\multicolumn{1}{c|}{} &\multicolumn{6}{c|}{Dirichlet parameters} \\
\cline{2-7} \multicolumn{1}{c|}{} & $c_{1k}$ & $c_{2k}$ & $c_{3k}$ & $\hat{c}_{1k}$ & $\hat{c}_{2k}$ & $\hat{c}_{3k}$\\
\hline  $k=1$ & 15  & 15 &  1  & 14.97  & 14.85 &  1.00\\
\hline  $k=2$ & 1   & 8  &  8  & 1.05   & 8.24  &  8.19 \\
\hline  $k=3$ & 3   & 1  & 3   & 3.12    & 1.02   & 3.03\\
  \hline
\end{tabular}
\label{tab:Dirichlet_coeff}
\end{table}
Fig. \ref{fig:labels_NCM_SigmaLR_epsN_K3_UsNCM} (right) displays the estimated classification map obtained with the proposed algorithm. This map is in a very good agreement with the ground truth shown in Fig. \ref{fig:labels_NCM_SigmaLR_epsN_K3_UsNCM} (left). Note that the Dirichlet parameters used in this simulation correspond to three distinguishable classes that are well separated using the proposed algorithm. The obtained classification results can also be observed with the data projected in the plane associated with the two most discriminant principle components as shown in Fig. \ref{fig:Projection_Data_VCA_UsNCM_epsN_K3}. The proposed algorithm also allows the Dirichlet parameters to be estimated accurately as shown in Table \ref{tab:Dirichlet_coeff}.

\begin{figure}[h!]
\centering \subfigure[]{\includegraphics[width=0.49\figwidth]{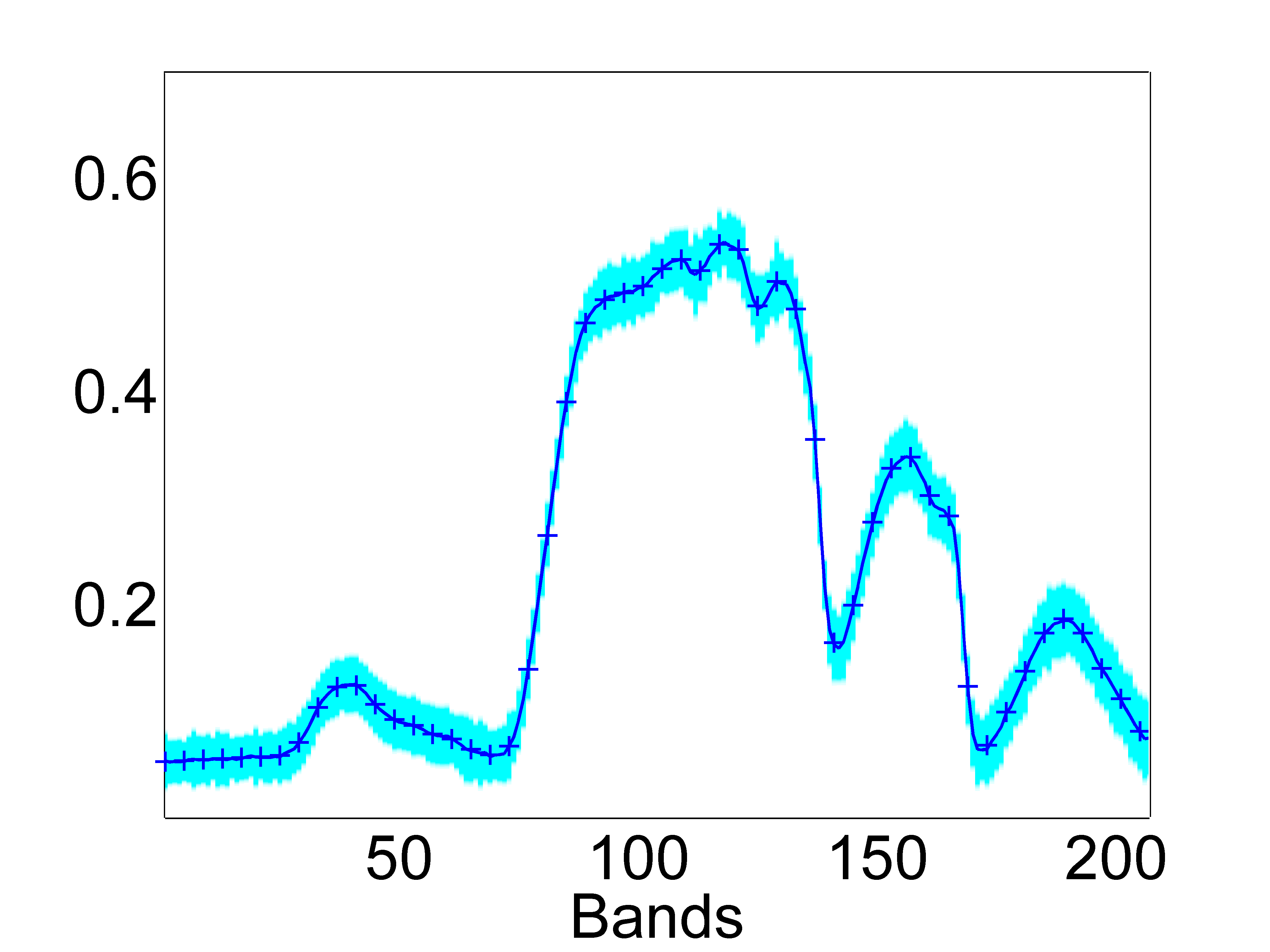}}
\subfigure[]{\includegraphics[width=0.49\figwidth]{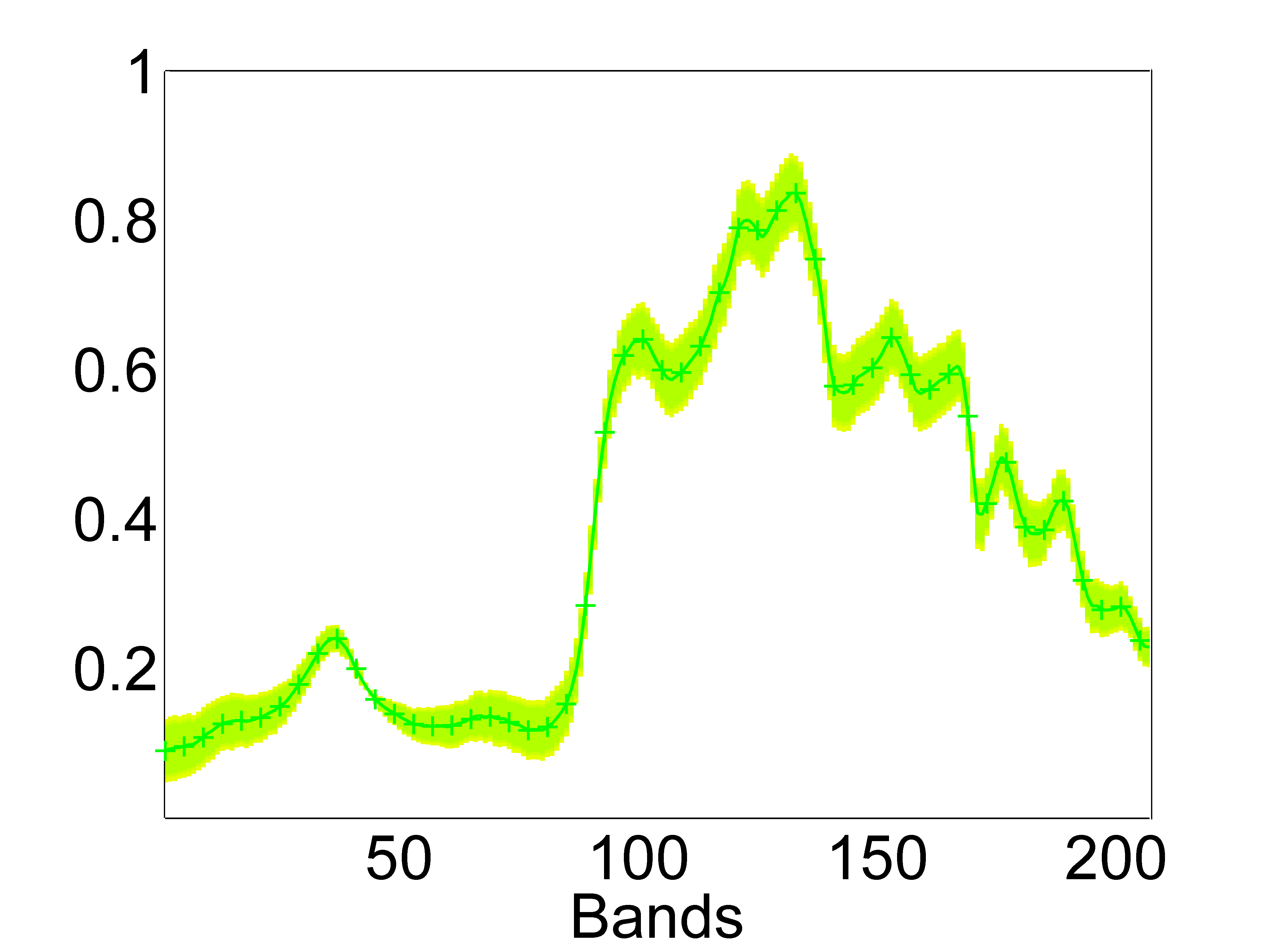}}
\subfigure[]{\includegraphics[width=0.49\figwidth]{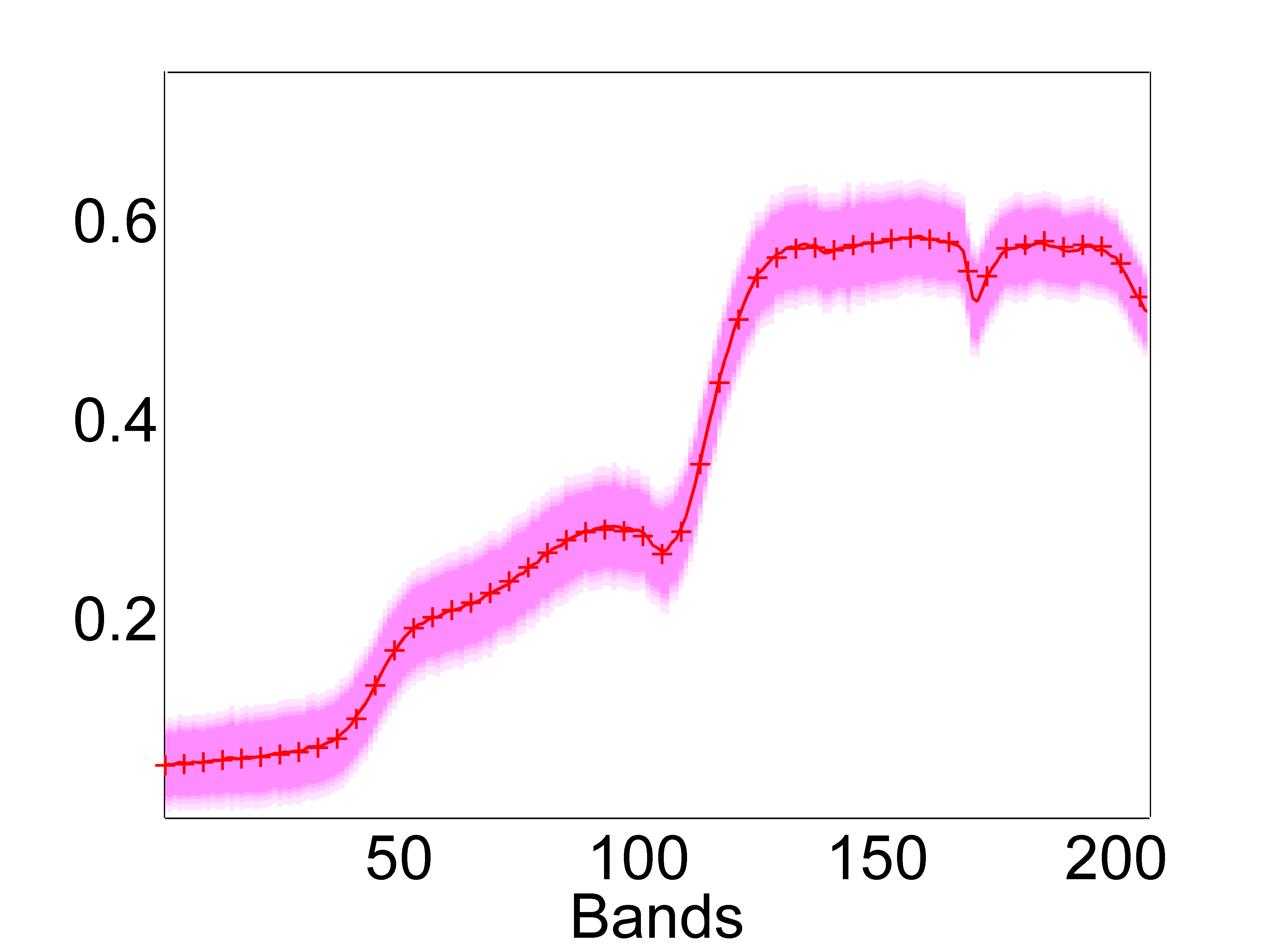}}
\subfigure[]{\includegraphics[width=0.49\figwidth]{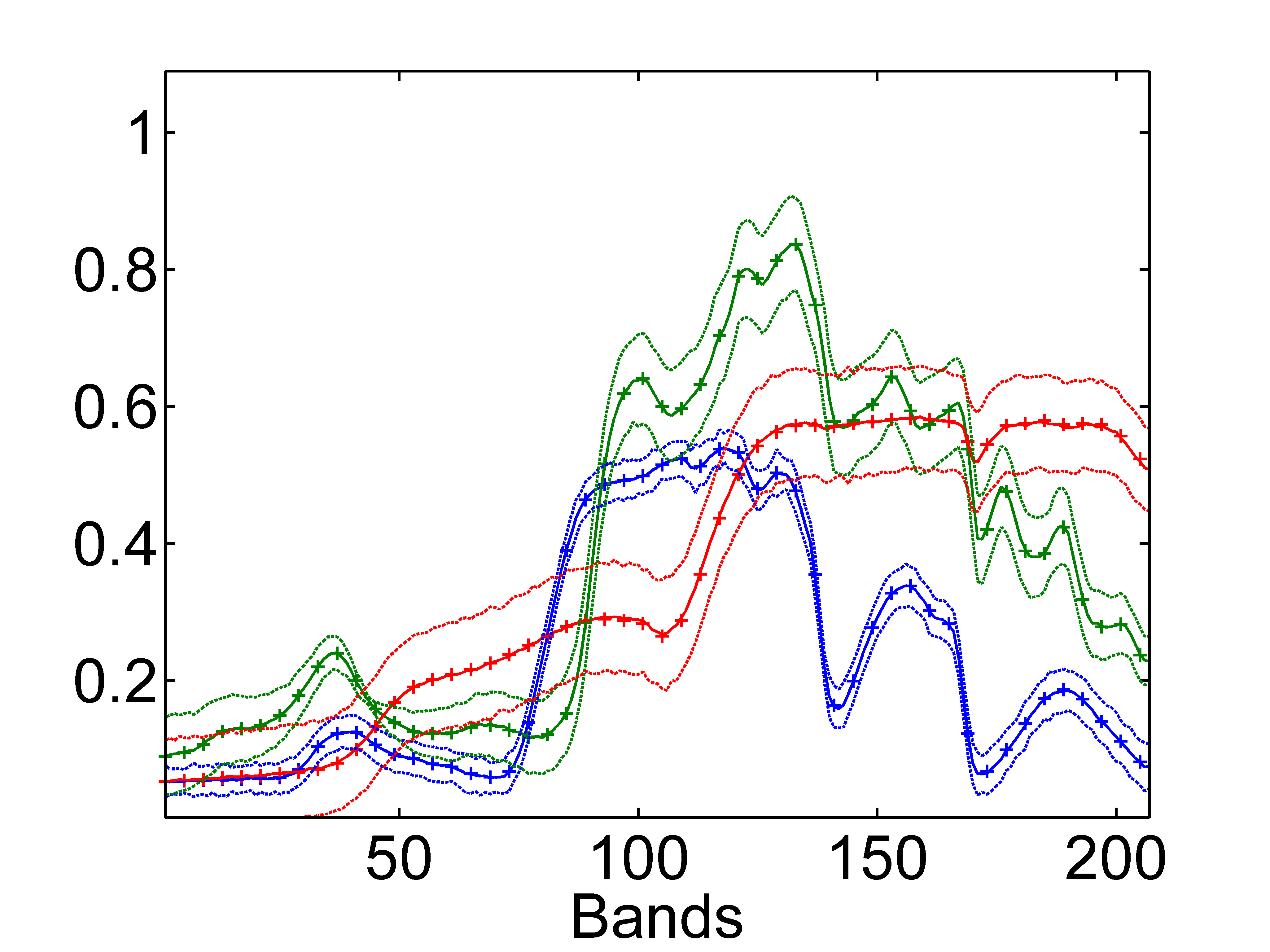}}
\caption{Actual endmembers (crosses) and endmember means estimated by UsGNCM (continuous lines). The estimated endmember distributions are represented in (a), (b), (c) by colored areas. The bottom-right figure (d) shows the endmembers estimated by UsGNCM $\pm 3 \sigma$ (dashed lines).} \label{fig:Endmembers_UsNCM_epsN_K3}
\end{figure}

A significant advantage of the proposed algorithm is its ability to estimate the endmember means and variances. Fig. \ref{fig:Projection_Data_VCA_UsNCM_epsN_K3}  shows the estimated endmembers obtained using the VCA algorithm (diamonds) \cite{Nascimento2005}, the UsLMM algorithm (circles) \cite{Dobigeon2009} and the proposed UsGNCM approach (triangles). Contrary to the VCA algorithm that provides bad endmember estimates because of the absence of pure pixels in the image, both UsLMM and UsGNCM strategies  yield good endmember estimations. As explained before, the good performance of the  UsGNCM algorithm can be explained by the fact that it is able mitigate the endmember variability. Fig. \ref{fig:Endmembers_UsNCM_epsN_K3} displays the endmember means (continuous lines), the endmember distributions (colored areas in Figs. \ref{fig:Endmembers_UsNCM_epsN_K3}(a), (b) and (c)) and the associated variability intervals defined by mean $\pm 3 \sigma$ (Fig. \ref{fig:Endmembers_UsNCM_epsN_K3} (d)). Fig. \ref{fig:Variation_variance_Endmembers_NCM_SigmaLR_K3} displays
the actual and estimated endmember variances for the three endmembers that are clearly in good agreement.
These results show the good performance of the proposed approach that fully exploits the spatial (segmentation map, abundances and noise variances) and spectral (endmember means and variances) correlations. The next section studies the robustness of the proposed approach with respect to the number of endmembers and the image size (number of pixels).

%\begin{figure}[h!]
%\centering
%\includegraphics[width=0.75\figwidth]{images/Endmembers_UsNCM_epsN_K3.png}
%\caption{Actual endmembers (crosses), endmembers estimated by UsGNCM (continuous lines) and  endmembers estimated by UsGNCM $\pm 3 \sigma$ (dashed lines).} \label{fig:Endmembers_UsNCM_epsN_K3}
%\end{figure}

%\begin{figure}[h!]
%\centering \subfigure[]{\includegraphics[width=0.32\figwidth,height=7.5cm]{images/Endmembers_VCA_UsNCM_epsN_K3_1}}
%\subfigure[]{\includegraphics[width=0.32\figwidth,height=7.5cm]{images/Endmembers_VCA_UsNCM_epsN_K3_2}}
%\subfigure[]{\includegraphics[width=0.32\figwidth,height=7.5cm]{images/Endmembers_VCA_UsNCM_epsN_K3_3}}
%\caption{Actual endmember (red line), endmembers estimated by VCA (black line), endmembers estimated by UsLMM (cyan circle) and  endmembers estimated by UsNCM $\pm 3 \sigma$ (dashed and continuous blue lines).} \label{fig:Endmembers_VCA_UsLMM_UsNCM}
%\end{figure}

\subsection{Performance as a function of the number of endmembers and the image size} \label{subsec:Performance_with_respect_to_the_number_of_endmembers_and_pixels}
\begin{figure}[h!]
\centering
\includegraphics[width=0.85\figwidth]{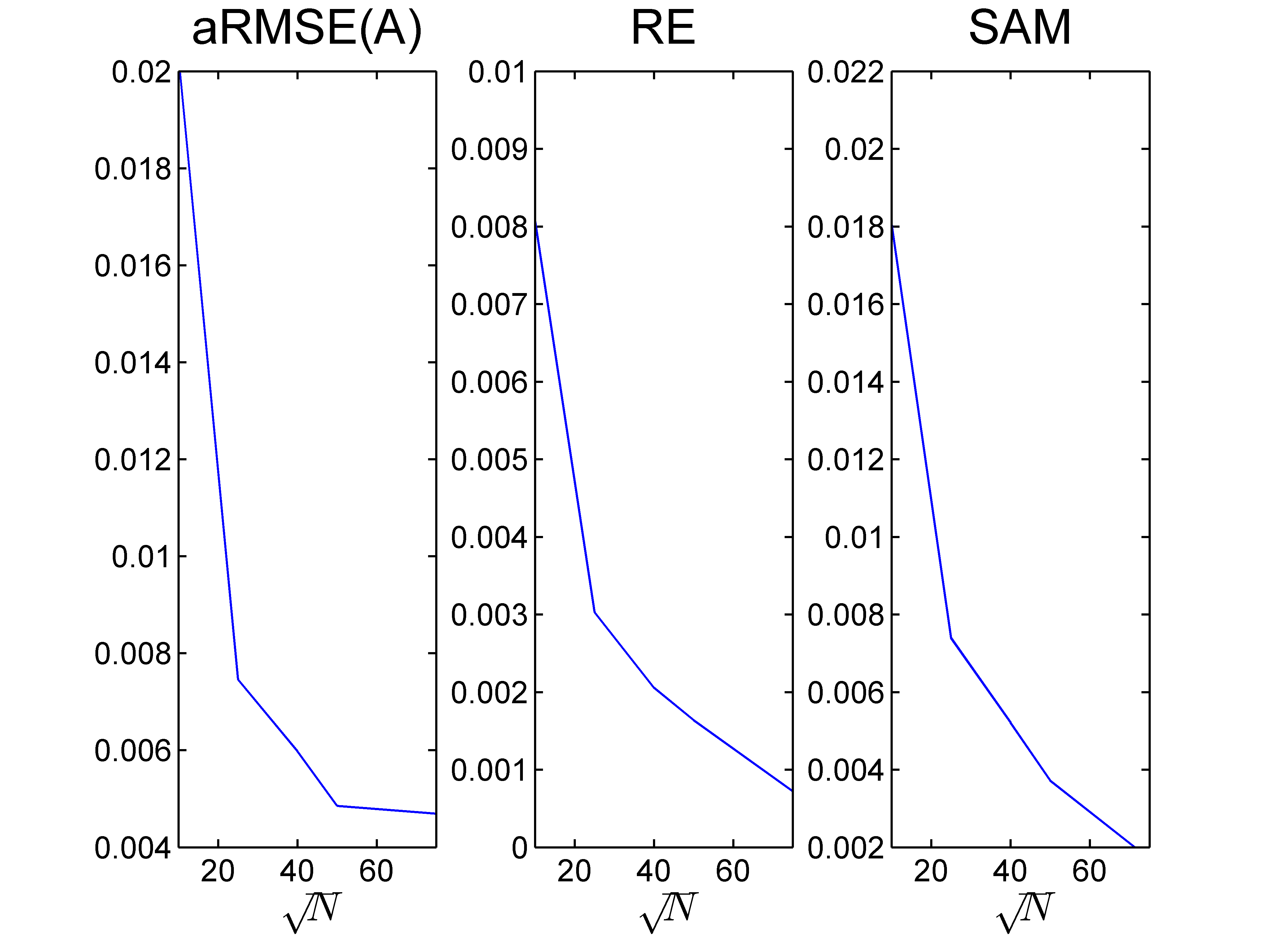}
\caption{UsGNCM performance for different numbers of pixels.} \label{fig:NCM_SigmaLR_epsN_varN_RMSE_RE_SAM}
\end{figure}

The UsGNCM algorithm estimates many parameters which might require a lot of observations in order to obtain acceptable performance. The first part of this section deals with this problem by analyzing the proposed algorithm when varying the number of observed pixels. The considered image has been generated using the three endmembers considered in Section \ref{subsec:Performance_of_the_proposed_algorithm} with abundances uniformly distributed in the simplex $\calS$ defined by the positivity and sum-to-one constraints (the corresponding Dirichlet parameters are $c_{rk} = 1, \forall r, \forall k$).
Fig. \ref{fig:NCM_SigmaLR_epsN_varN_RMSE_RE_SAM} shows the obtained aRMSE$\left(\bsA\right)$, RE and SAM when varying the size of the observed image. As expected, the unmixing performance improves by increasing the number of observations. This figure also shows that the aRMSE$\left(\bsA\right)$ converges to a constant value for $\sqrt{N}>50$ while RE and SAM continue to improve when increasing $N$. Note, however, that the obtained results are quite good for $N\geq100$.
\indent The second part of this section analyzes the behavior of UsGNCM with respect to the number of endmembers. Table \ref{tab:Variation_R} shows the obtained aRMSE$\left(\bsA\right)$, aRMSE$\left(\bsM\right)$ and aSAM$\left(\bsM\right)$ criteria for $R=\left\lbrace 3, 4, 5, 6\right\rbrace$. The considered endmembers are construction concrete, green grass, micaceous loam, olive green paint, bare red brick, and galvanized steel metal. These spectra
have been extracted from the spectral libraries provided with the ENVI software \cite{ENVImanual2003}. As previously, the images associated with $R=\left\lbrace 3, 4, 5, 6\right\rbrace$ have been generated with abundances uniformly distributed in the simplex $\calS$. As expected, increasing the number of parameters (i.e., increasing $R$) reduces the estimation performance. However, the obtained results are still acceptable confirming the robustness of UsGNCM with respect to the number of endmembers  $R$.
\renewcommand{\arraystretch}{1.1}
\begin{table}[h] \centering
\centering \caption{UsGNCM performance for different number of endmembers.}
\begin{tabular}{|c|c|c|c|}
  \cline{2-4}
\multicolumn{1}{c|}{} & aRMSE$\left(\bsA\right)$ & aRMSE$\left(\bsM\right)$& aSAM$\left(\bsM\right)$ \\
\multicolumn{1}{c|}{} & $(\times 10^{-2})$ & $(\times 10^{-2})$ & $(\times 10^{-2})$ \\
\hline  $R=3$ & 0.5734  & 0.1826 &  0.4306 \\
\hline  $R=4$ & 0.542   & 0.2115 &  0.5073 \\
\hline  $R=5$ & 0.8053  & 0.2790 &  0.6837 \\
\hline  $R=6$ & 1.4049  & 0.8404 &  1.6550 \\
  \hline
\end{tabular}
\label{tab:Variation_R}
\end{table}

%\clearpage
\subsection{Comparison with state-of-the-art algorithms} \label{subsec:Comparison_with_other_algorithms}
This section evaluates the performance of the proposed UsGNCM algorithm for different images. All images  have been constructed using $R=3$ endmembers with truncated abundances (with $a_i <0.9$, $\forall i \in {1,\cdots,R}$) to avoid the presence of pure pixels. The remaining parameters have been defined as follows
\begin{itemize}
\item the image $I_1$ has been generated according to the GNCM model with $K=1$ class and abundances uniformly distributed in the simplex $\calS$. The endmember variances have been adjusted as in Fig. \ref{fig:Variation_variance_Endmembers_NCM_SigmaLR_K3}. The noise variance is $\psi_{n}^2 =10^{-7}$.
\item the image $I_2$ is the GNCM image used in Section \ref{subsec:Performance_of_the_proposed_algorithm}.
\item the image $I_3$ has been generated according to the LMM model with $K=3$ classes and the Dirichlet parameters of Table \ref{tab:Dirichlet_coeff}. The noise variances vary linearly with respect to the spectral bands with $$\psi^2_l  = 10^{-4} \left(\frac{4}{L-1} l + \frac{L+3}{L-1} \right),     \textrm{  for  } l\in [1,\cdots,L].$$
\end{itemize}
These images are processed using different unmixing strategies that are compared to the proposed UsGNCM algorithm. More precisely, we have considered the following unmixing algorithms
\begin{itemize}
\item VCA+FCLS: the endmembers are extracted from the whole image using VCA and  the abundances are estimated using the FCLS algorithm \cite{Heinz2001}.
\item UsLMM: the unsupervised Bayesian algorithm of \cite{Dobigeon2009} is used to jointly estimate the endmembers and abundances.
\item AEB: this is the automated endmember bundles algorithm proposed in \cite{Somers2012}. We consider a $10\%$ image subset and the VCA algorithm to extract the endmembers. For each pixel, the $3$ endmembers that provide the smallest RE are selected.
\item UsNCM: the proposed unmixing strategy with $\psi_n=0$ (i.e., the additive noise $\bse_n$ of \eqref{eqt:Normal_compositional_model1} is removed). Note that the resulting algorithm reduces to the NCM model.
\end{itemize}
The first two algorithms provide one estimate for each endmember while the other algorithms estimate endmember variability. Note that the UsNCM is introduced to study the effect of the additive noise. Table \ref{tab:Results_Synth} reports the quality of the estimated abundances and endmembers by unmixing the three images with the different algorithms. This table shows bad performance for VCA+FCLS and AEB algorithms which is mainly due to the absence of pure pixels in the considered images. The UsLMM provides good results for the three images. However, it appears to be sensitive to the variation of endmember/noise variances with respect to the spectral band and to the spatial correlations between adjacent pixels. Indeed, the UsLMM did not consider spatial correlation which leads to a performance reduction when processing the images $I_2$ and $I_3$. Note also that the UsLMM algorithm provides one estimate for each endmember and does not take into account the spatial variability of endmembers in the processed images. The best performance is generally obtained by the proposed UsNCM and UsGNCM strategies that provide almost similar results. However, the UsGNCM algorithm is more robust than UsNCM  when processing the LMM image $I_3$. Moreover, the UsGNCM provides the best endmember estimates as highlighted by the criteria ARE and ASAM. These results confirm the superiority of the proposed approach in presence of endmember variability, spatial correlation between pixels and in absence of pure pixels in the observed scene.

%VCA+FCLS, AEB
%Bad results because of absence of pure pixels
%
%UsLMM robust for all images,
%- no endmember variability
%- variance variable
%
%UsGNCM>UsNMC
%UsGNCM robust for LMM
%better endmember

%\begin{itemize}
%\item $I_1$: generated according to NCM with $a_i <0.9$, $\forall i \in {1,\cdots,3}$, $K=1$, abundances uniformly distributed  in the simplex and $\sigma^2_{rl}$ varying for $r$ and $l$
%\item $I_2$: generated according to NCM with $a_i <0.9$, $\forall i \in {1,\cdots,3}$, $K=3$, abundances distributed in three classes in the simplex and $\sigma^2_{rl}$ varying for $r$ and $l$
%\item $I_3$: generated according to LMM with $a_i <0.9$, $\forall i \in {1,\cdots,3}$, $K=3$, abundances distributed in three classes in the simplex and $\kappa^2_l  = 10^{-4} \left(\frac{4}{L-1} l + \frac{L+3}{L-1} \right)$
%\end{itemize}

\renewcommand{\arraystretch}{1.1}
\begin{table}[h] \centering
\centering \caption{Results on synthetic data.}
\begin{tabular}{|c|c|c|c|c|c|c|c|c|c|c|}
\cline{3-11}
\multicolumn{2}{c|}{} &  \multicolumn{9}{c|}{Criteria $(\times 10^{-2})$} \\
\cline{3-11}
\multicolumn{2}{c|}{}                        &  aRMSE   & RMSE   & RMSE   & RMSE   & SAM   & SAM   & SAM   &  aRMSE   &   aSAM \\
\multicolumn{2}{c|}{}                        &  $\left(\bsA\right)$   & $\left(\bsm_1\right)$   & $\left(\bsm_2\right)$   & $\left(\bsm_3\right)$   & $\left(\bsm_1\right)$   & $\left(\bsm_2\right)$   & $\left(\bsm_3\right)$   &  $\left(\bsM\right)$   &   $\left(\bsM\right)$ \\
\hline \multirow{3}{*}{image $I_1$} & VCA+FCLS & 4.78 &  2.29 &  1.97 &  2.31 &  6.14 &  1.71 &  5.74 &  2.20 &  4.53 \\
\cline{2-11}                        & UsLMM    & 0.52 &  0.25 &  0.10 &  \textbf{0.15} &  0.77 &  0.20 &  \textbf{0.33} &  0.18 &  0.43 \\
\cline{2-11}\multirow{2}{*}{(GNCM,} & AEB      & 3.73 &  2.33 &  1.80 & 2.55  &  4.92 &  2.04 &  7.75 &  2.25 &  4.90 \\
\cline{2-11}\multirow{2}{*}{$K=1$)} & UsNCM    & \textbf{0.48} &  0.09 &  0.10 &  0.21 &  0.30 &  0.20 &  0.43 &  0.14 &  0.31 \\
\cline{2-11}                        & UsGNCM   & \textbf{0.48} &  \textbf{0.07} &  \textbf{0.09} &  0.18 &  \textbf{0.25} &  \textbf{0.19} &  0.40 &  \textbf{0.12} &  \textbf{0.28} \\
\hline
\hline \multirow{3}{*}{image $I_2$} & VCA+FCLS & 3.71  & 2.89  & 2.98  & 2.08  &  9.54  &  5.60  &  5.09   &  2.68 &   6.74 \\
\cline{2-11}                        & UsLMM    & 0.76  & 0.21  & 0.40  & 0.73  &  0.67  &  0.86  &  1.28   &  0.49 &   0.94 \\
\cline{2-11}\multirow{2}{*}{(GNCM,} & AEB      & 9.46  & 3.48  & 4.67  & 4.37  &  7.96  & 13.26  &  4.94   &  4.20 &   8.72 \\
\cline{2-11}\multirow{2}{*}{$K=3$)} & UsNCM    & 0.56  & \textbf{0.14}  & \textbf{0.11}  & 0.28  &  \textbf{0.39}  & \textbf{0.21}  &  0.69   &  0.19 &   0.43 \\
\cline{2-11}                        & UsGNCM   & \textbf{0.48}  & \textbf{0.14}  & \textbf{0.11}  & \textbf{0.21}  &  0.46  &  0.26  &  \textbf{0.51}   &  \textbf{0.16} &   \textbf{0.41} \\
\hline
\hline \multirow{3}{*}{image $I_3$} & VCA+FCLS & 9.51 &  3.63 &  6.21 &  2.61 & 12.20 &  7.73 &  5.60 &  4.42 &  8.51 \\
\cline{2-11}                        & UsLMM    & 1.01 &  0.75 &  \textbf{0.18} &  \textbf{0.34} &  2.62 &  \textbf{0.30} &  \textbf{0.73} &  0.49 &  1.22 \\
\cline{2-11}\multirow{2}{*}{(LMM,}  & AEB      & 9.30 &  5.98 &  4.67 &  4.61 & 16.05 &  5.86 & 10.84 &  5.13 & 10.92\\
\cline{2-11}\multirow{2}{*}{$K=3$)} & UsNCM    & 0.86 &  0.45 &  0.44 &  0.55 &  1.55 &  0.91 &  0.99 &  0.48 &  1.15 \\
\cline{2-11}                        & UsGNCM   & \textbf{0.74} &  \textbf{0.20} &  0.29 &  0.47 &  \textbf{0.70} &  0.56 &  0.98 &  \textbf{0.34} &  \textbf{0.74} \\
  \hline
\end{tabular}
\label{tab:Results_Synth}
\end{table}

\section{Simulation results on real data} \label{sec:Simulation_results_on_real_data}

\subsection{Description of the Hyperspectral Data} \label{subsec:Description_of_the_Hyperspectral_Data}
\begin{figure}[h!]
\centering
\includegraphics[width=1\figwidth]{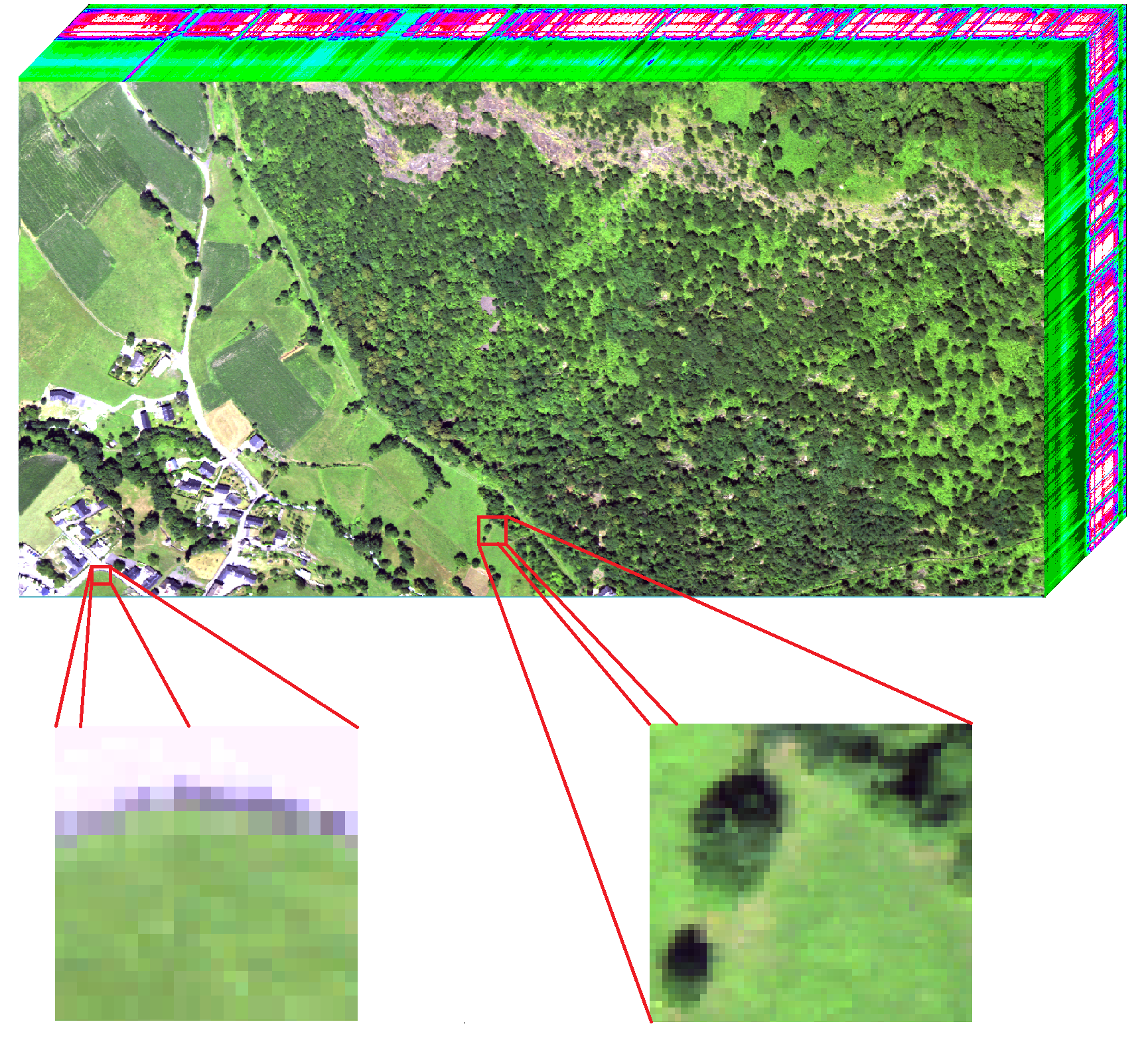}
\caption{Real Madonna image and the considered subimages shown in true colors. (Right) scene 1, (left) scene 2 } \label{fig:Madonna_scene}
\end{figure}

This section illustrates the performance of the proposed UsGNCM algorithm when applied to a real hyperspectral data set. The
real image used in this section was acquired in $2010$ by the Hyspex hyperspectral scanner over Villelongue, France (00 03'W and 4257'N). The dataset contains $L=160$ spectral bands recorded from the visible to near infrared with a spatial resolution of $0.5$ m  \cite{Sheeren2011}. The proposed unmixing algorithm has been applied to two subimages: scene  $\#1$ of size $50 \times 50$ which is composed of $R=4$ components: tree, grass, soil and shadow (see Fig. \ref{fig:Madonna_scene} (right)), and scene $\#2$ of size $31 \times 31$ which is composed of $R=3$ components: grass, road and ditch (see Fig. \ref{fig:Madonna_scene} (left)).

\subsection{Endmember  Variability} \label{subsec:Endmembers_variability}
\begin{figure}[h!]
\centering
\includegraphics[width=1\figwidth]{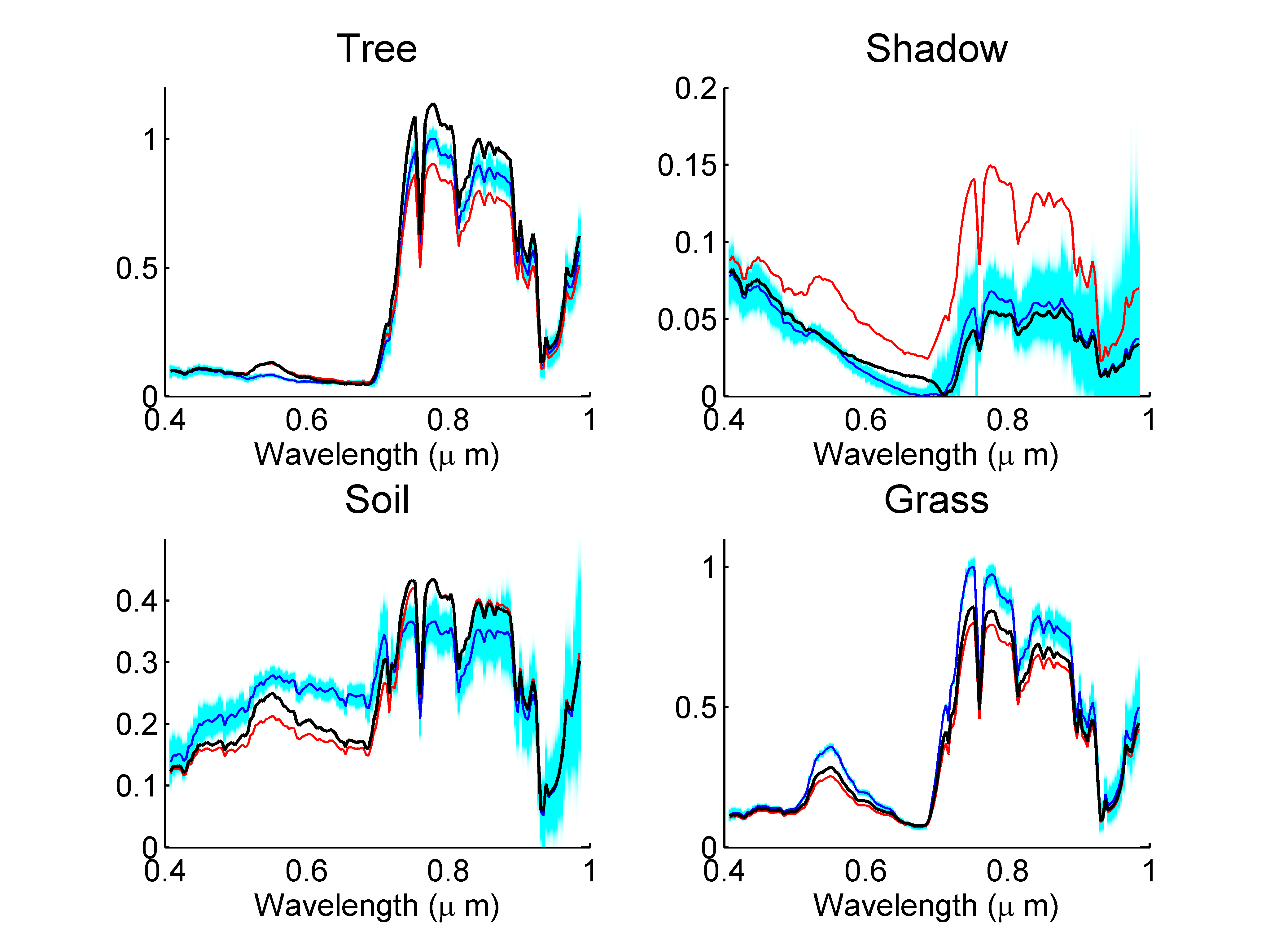}
\caption{The $R = 4$ endmembers estimated by VCA (continuous red lines), UsLMM (continuous black lines), UsGNCM (continuous blue lines) and the estimated endmember distribution (blue level areas) for the Madonna image.  } \label{fig:Endmembers_Madona_VCA_UsLMM_UsNCM_epsN_R4_K4}
\end{figure}

\begin{figure}[h!]
\centering
\includegraphics[width=1\figwidth]{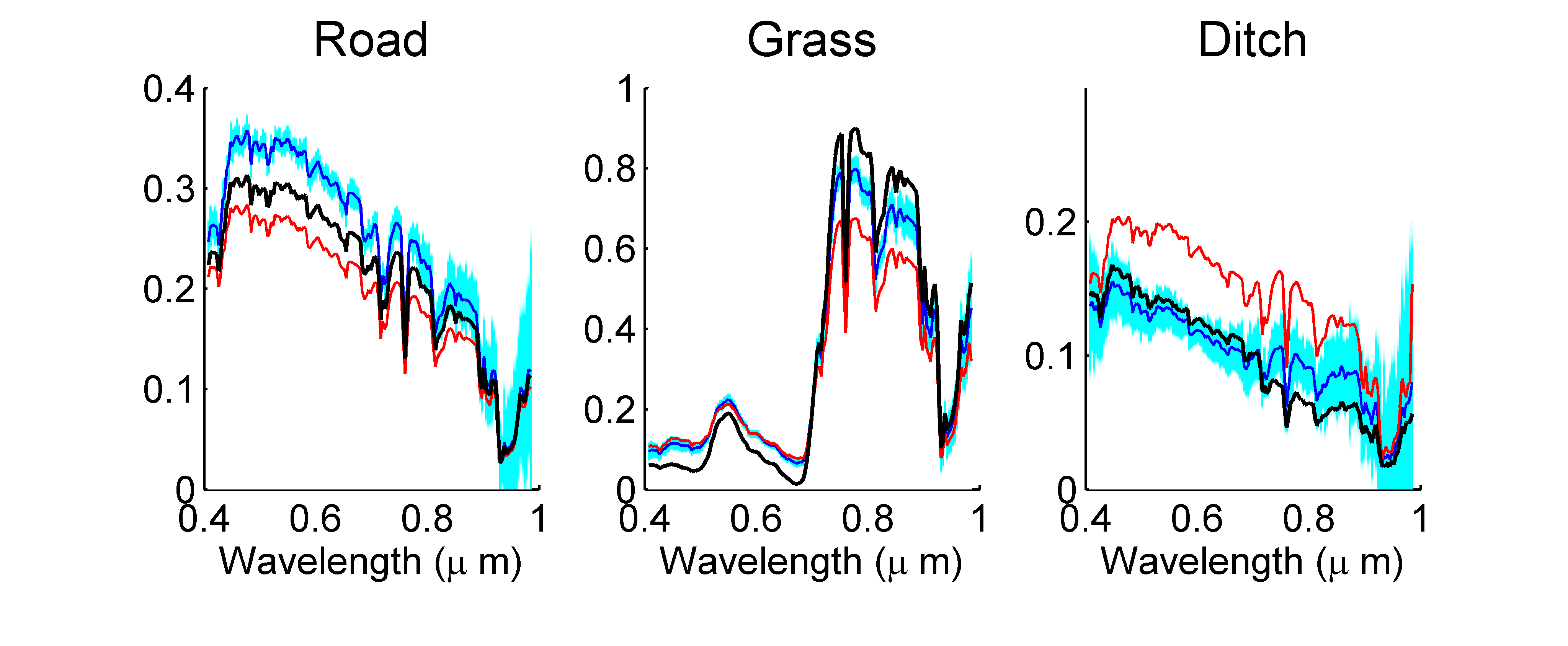}
\caption{The $R = 3$ endmembers estimated by VCA (continuous red lines), UsLMM (continuous black lines), UsGNCM (continuous blue lines) and the estimated endmember distribution (blue level areas) for the Madonna image.  } \label{fig:Endmembers_Madona_VCA_UsLMM_UsNCM_epsN_R4_K4_scene2}
\end{figure}

The proposed UsGNCM algorithm can estimate both the endmember means and variances. Fig. \ref{fig:Endmembers_Madona_VCA_UsLMM_UsNCM_epsN_R4_K4} compares the endmember estimates of this algorithm  with those obtained with VCA and UsLMM when considering scene $\#1$. The estimated endmembers are globally in good agreement. Note that VCA provides a different shadow endmember because it estimates the endmember as the purest pixel in the image while UsLMM and UsGNCM estimate both the abundances and endmembers resulting in a better shadow estimate (lower amplitude). Moreover, the proposed algorithm provides endmember distributions (blue level areas in Fig. \ref{fig:Endmembers_Madona_VCA_UsLMM_UsNCM_epsN_R4_K4}) which measure the endmember variability in the considered image. It can be seen that the higher relative variation is obtained for the shadow spectrum because of its low  amplitude. Moreover, the variation is more pronounced for high spectral bands ($l>80$) which is in agreement with the results presented in \cite{Altmann2014b}. Fig. \ref{fig:Endmembers_Madona_VCA_UsLMM_UsNCM_epsN_R4_K4_scene2} shows the obtained endmembers when considering scene $\#2$. This figure presents similar results between UsGNCM and UsLMM, especially for capturing spectral components having low amplitudes as for ditch.

\subsection{Abundance Estimation and Image Classification} \label{subsec:Abundance_Estimation_and_image_classification}
\begin{figure}[h!]
\centering
\includegraphics[width=0.85\figwidth]{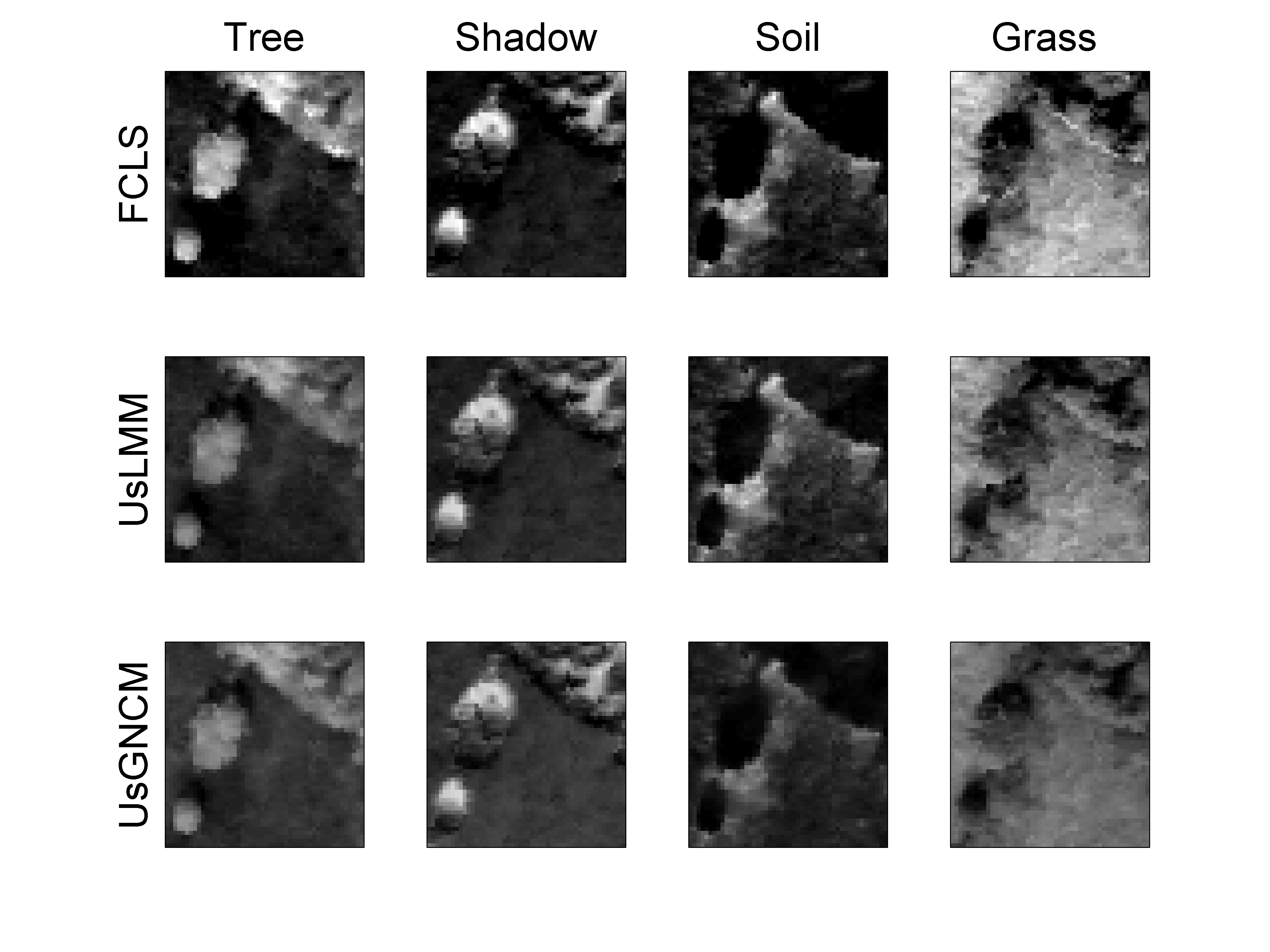}
\caption{Abundance maps estimated by FCLS (top), UsLMM (middle) and the proposed UsGNCM (bottom) for the Madonna image.} \label{fig:Abundances_Madona_VCA_UsLMM_UsNCM_epsN_R4_K4}
\end{figure}

\begin{figure}[h!]
\centering \subfigure[Classification map.] {\includegraphics[width=0.45\figwidth]{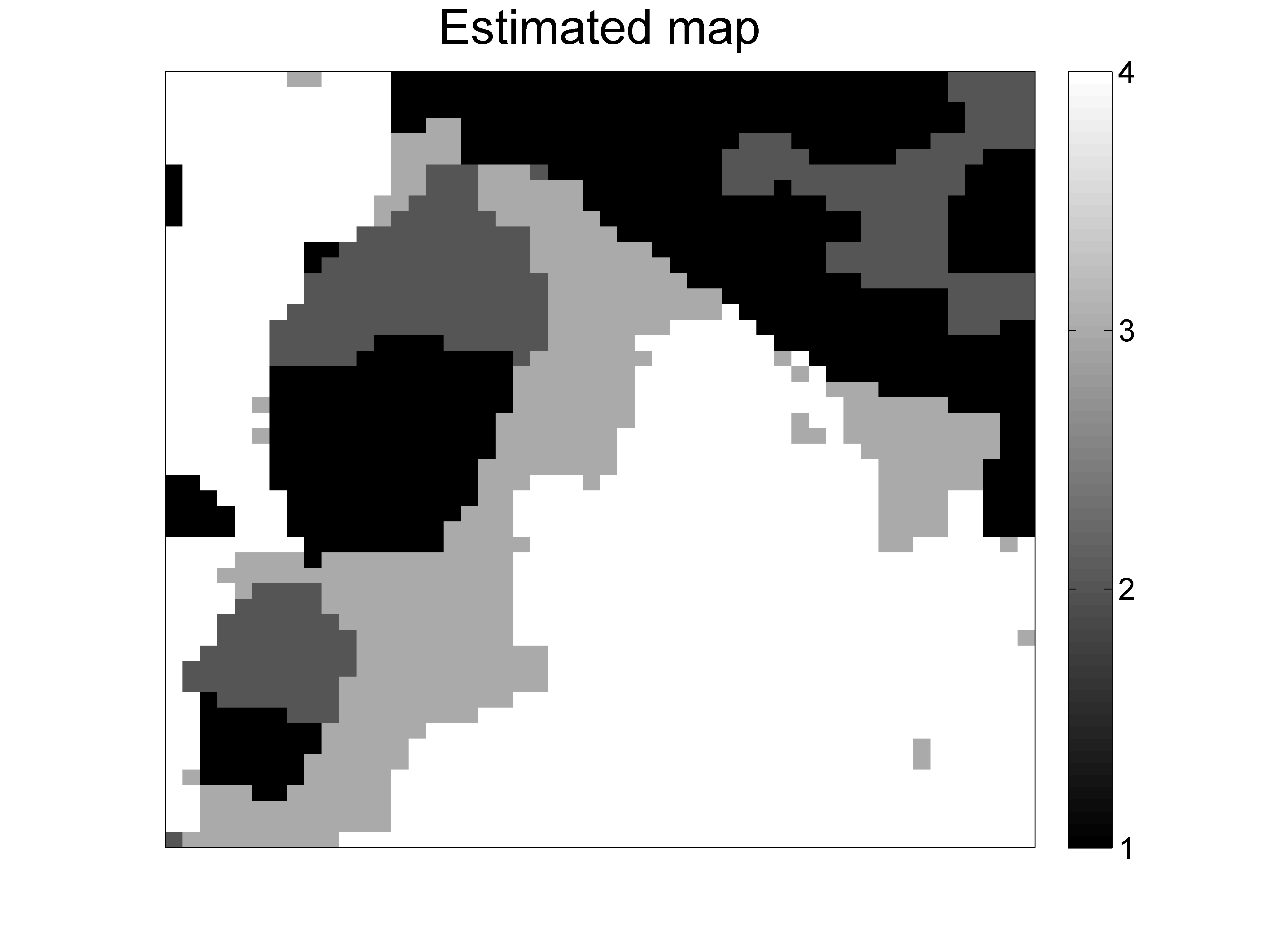}}
\subfigure[Noise variances.] {\includegraphics[width=0.45\figwidth]{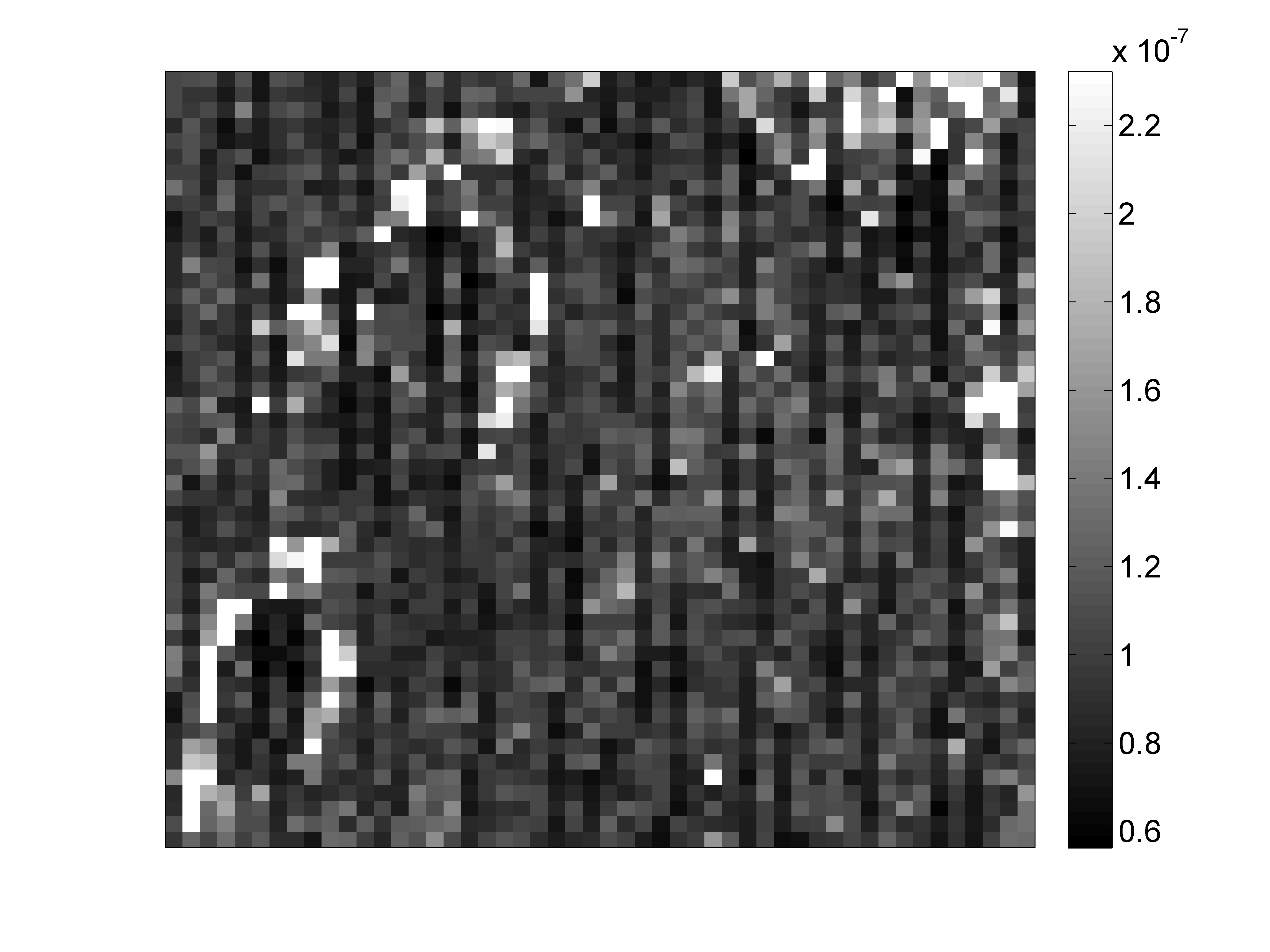}}
\caption{Estimated maps with the UsGNCM algorithm for the scene $\#1$ of Madonna image. (a) Classification map and (b) noise variances. } \label{fig:labels_noise_Madona_arbre_UsNCM_epsN}
\end{figure}

\begin{figure}[h!]
\centering \subfigure[Classification map.] {\includegraphics[width=0.45\figwidth]{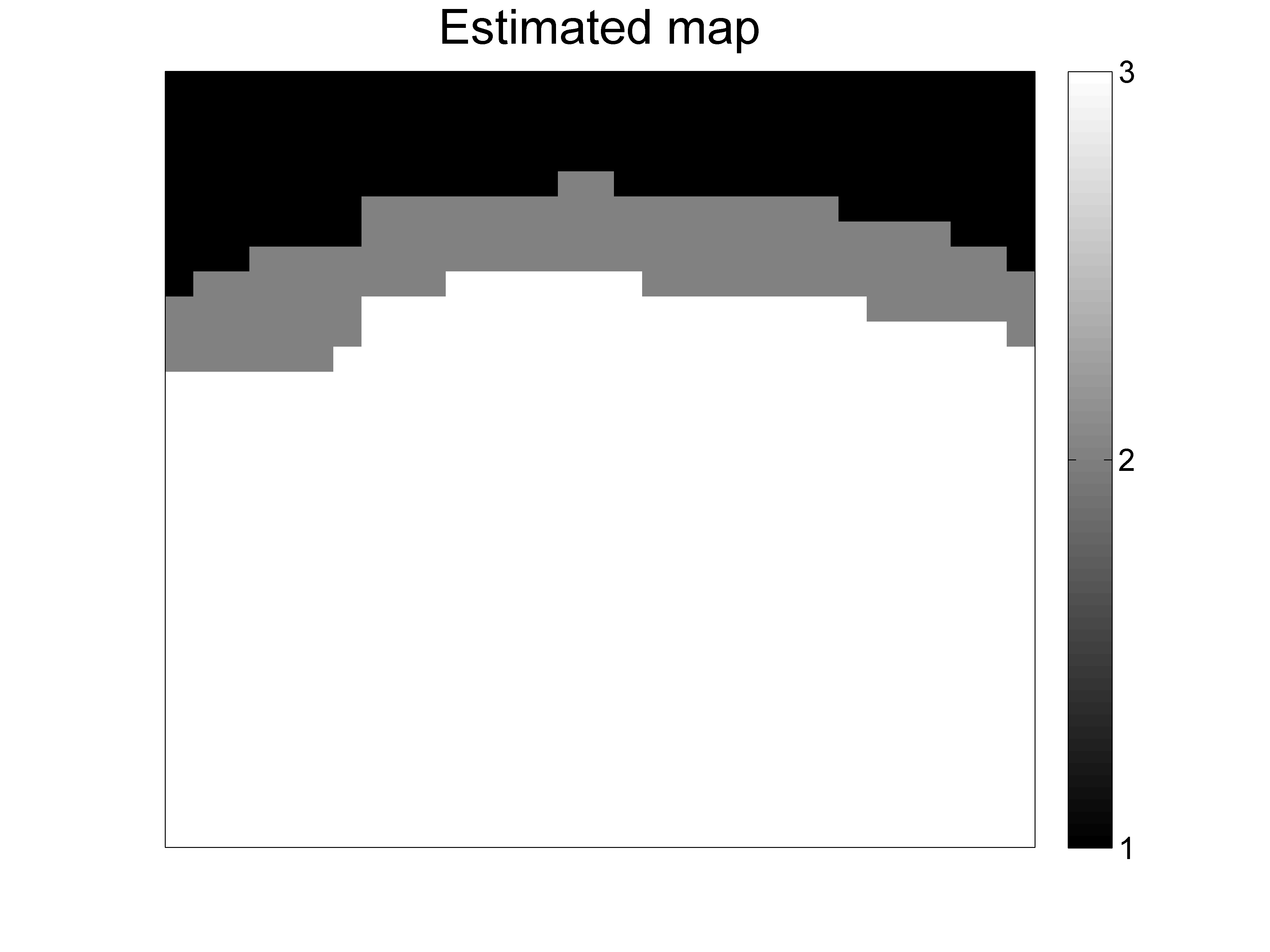}}
\subfigure[Noise variances.] {\includegraphics[width=0.45\figwidth]{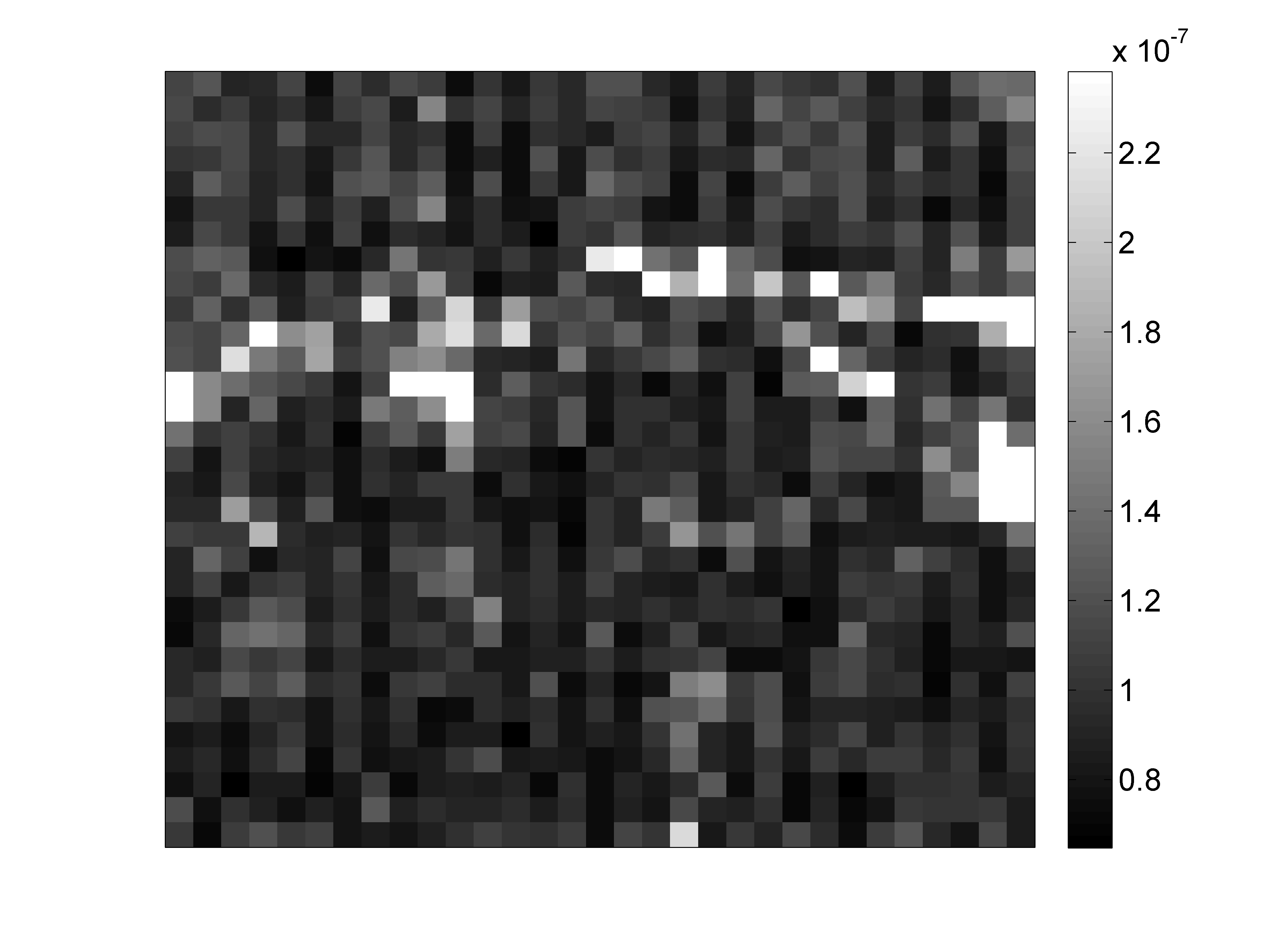}}
\caption{Estimated maps with the UsGNCM algorithm for the scene $\#2$ of Madonna image. (a) Classification map and (b) noise variances. } \label{fig:labels_noise_Madona_arbre_UsNCM_epsN_scene2}
\end{figure}

The fraction maps of  scene $\#1$ estimated by the proposed method are shown in Fig. \ref{fig:Abundances_Madona_VCA_UsLMM_UsNCM_epsN_R4_K4} (bottom). Note that a white (black) pixel indicates a large (small) proportion of the corresponding materials. These pictures are in good agreement with the FCLS and UsLMM results shown in Fig. \ref{fig:Abundances_Madona_VCA_UsLMM_UsNCM_epsN_R4_K4} (top)  and (middle), respectively. Note that the compared algorithms also provide similar abundance maps when considering scene $\#2$. However, these results are not presented here for brevity. In addition to unmixing, UsGNCM also provides a spatial segmentation of the considered scenes as shown in Fig. \ref{fig:labels_noise_Madona_arbre_UsNCM_epsN}(a) for scene $\#1$ and Fig. \ref{fig:labels_noise_Madona_arbre_UsNCM_epsN_scene2}(a) for scene $\#2$. These classifications clearly highlight the area of each physical element in the scene. Indeed, for scene $\#1$ we have $4$ classes that represent tree, shadow, soil and grass zones while for scene $\#2$ we have $3$ classes representing road, ditch and grass areas. Table \ref{tab:Dirichlet_coeff_Real} finally reports the estimated Dirichlet parameters and the number of pixels for each spatial class when considering scene $\#1$. These parameters suggest a highly non uniform distribution over the simplex which promote the use of the proposed approach.

\renewcommand{\arraystretch}{1.1}
\begin{table}[h] \centering
\centering \caption{Estimated Dirichlet parameters for the Madonna image.}
\begin{tabular}{|c|c|c|c|c|c|}
  \cline{2-6}
\multicolumn{1}{c|}{} &\multicolumn{4}{c|}{Dirichlet parameters} & number of \\
\cline{2-5} \multicolumn{1}{c|}{} & $\hat{c}_{1k}$ & $\hat{c}_{2k}$ & $\hat{c}_{3k}$ & $\hat{c}_{4k}$ & pixels\\
\hline  $k=1$ & 7.8767  & 2.8933   &  1.0139  &  5.1277   & 613\\
\hline  $k=2$ & 2.8914  & 7.6524   &  1.3115  &  1.7289   & 318\\
\hline  $k=3$ & 12.3176 &  16.1875 &  21.2009 &  21.1454  & 445\\
\hline  $k=4$ & 25.7654 &  26.4822 &  17.0927 &  49.9141  & 1124\\
  \hline
\end{tabular}
\label{tab:Dirichlet_coeff_Real}
\end{table}

\subsection{Residual Components} \label{subsec:Residual_components}
The proposed algorithm also provides a measure of the noise variance for each observed pixel. This parameter brings an information about pixels that are inaccurately described by a linear formulation, i.e., allows modeling errors to be quantified. Fig. \ref{fig:labels_noise_Madona_arbre_UsNCM_epsN}(b) shows the obtained noise variances for the scene $\#1$. This figure shows a higher error in the shadow area and around trees, i.e., for regions where possible interactions between physical components might occur (e.g., tree/soil) resulting in a more complex model than the proposed linear one. The noise variances associated with the scene $\#2$ are shown in Fig. \ref{fig:labels_noise_Madona_arbre_UsNCM_epsN_scene2}(b). This figure shows a higher error near the ditch area which might be due to the presence of nonlinearities as explained in  \cite{Altmann2014b}. Note finally that both Fig. \ref{fig:labels_noise_Madona_arbre_UsNCM_epsN}(b)  and Fig. \ref{fig:labels_noise_Madona_arbre_UsNCM_epsN_scene2}(b)  highlight the presence of regular vertical patterns that have also been observed in \cite{Fevotte2014} and were associated with a sensor defect or other miscalibration problems.

%Note finally that more real data results are provided in \cite{HalimiTR2014_EV}.

%%%%%%%%%%%%%%%%%%%%%%%%%%%%%%%%%%%%%%%%%%%%%%%%%%%
%%%%%%%%%%%%%%%%%%%%%%%%%%%%%%%%%%%%%%%%%%%%%%%%%%%
%%%%%%%%%%%%%%%%%%%%%%%%%%%%%%%%%%%%%%%%%%%%%%%%%%%
%\clearpage
\section{Conclusions} \label{sec:Conclusions}
This paper introduced a Bayesian model for unsupervised unmixing of hyperspectral images accounting for spectral variability. The proposed algorithm was based on a generalization of the normal compositional model and includes an additive Gaussian noise for modeling errors. This algorithm estimated the endmembers of the scene, their variabilities provided by their variances and the corresponding abundances. The observed image was also spatially segmented into regions sharing homogeneous abundance characteristics. The physical constraints of the abundances were ensured by choosing a Dirichlet distribution for each spatial class of the image. Due to the complexity of the resulting joint posterior distribution, a Markov chain Monte Carlo procedure based on a Gibbs algorithm was used to sample the posterior of interest and to approximate the Bayesian estimators of the unknown parameters using the generated samples. The sampling was achieved using an Hamiltonian Monte Carlo method which is well suited for problems with a large number of parameters. The proposed algorithm showed good performance when processing data presenting   endmember variability, spatial correlation between
pixels and in absence of pure pixels in the observed scene. UsGNCM fully exploits both the spatial dimension (segmentation, abundance and noise estimation) and the spectral dimension
(estimation of endmember means and variances).
Future work includes the study of endmember variability with nonlinear mixing models. This point is an interesting issue that is currently under investigation.

\newpage
\appendix[Derivatives of the potential functions] \label{app:Derivatives_of_the_potential_functions}
The derivative of $U$ with respect to $\bst_n$ is given by
\begin{equation}
\frac{\partial U}{\partial \bst_n}   = \frac{\partial U_1}{\partial \bsa_n} \frac{\partial \bsa_n}{\partial \bst_n}+
\frac{\partial U_2}{\partial \bst_n}+
\frac{\partial U_3}{\partial \bsa_n} \frac{\partial \bsa_n}{\partial \bst_n}
\label{eqt:dU_dZ}
\end{equation}
with
\begin{eqnarray}
\frac{\partial U_1}{\partial \bsa_n} & = & - \left[ \bLam_{:n} \odot \left(\bsy_{n}-\bsM \bsa_{n}\right)\right]^T \bsM + \frac{1}{2} \left[ \left(\bsy_{n}-\bsM \bsa_{n}\right) \odot \left(\bsy_{n}-\bsM \bsa_{n}\right)\right]^T \left(\frac{\partial \bLam_{:n}}{\partial \bsa_n}\right)^T \nonumber \\
\left(\frac{\partial \bLam_{ln}}{\partial \bsa_n}\right)^T & = & -2 \frac{\textrm{diag}(\bsa_n)   \bSig_{:,l} }{\bOme_{ln}^2} \nonumber \\
\frac{\partial U_3}{\partial \bsa_n} &= & \bsa_n^T \odot \left[ \bSig \bLam_{:n}  \right]^T \nonumber \\
\frac{\partial U_2}{\partial t_{rn}} &=&
-\frac{\sum_{i=r+1}^{R}{c_{ik}-1} }{t_{rn} } + \frac{c_{rk}-1}{1-t_{rn} }      , \; \forall r \in \left\lbrace 1,\cdots,R-1 \right\rbrace
 \nonumber \\
\label{eqt:dUi_dZ}
\end{eqnarray}
and
\begin{equation}
\frac{\partial a_{rn}}{\partial t_{in}} = \left\lbrace \begin{array}{ll}
0 & \textrm{if   } i>r \\
\frac{a_{rn}}{t_{in}-1} & \textrm{if   } i=r \\
\frac{a_{rn}}{t_{in}}   & \textrm{if   } i<r
\end{array}
\right..
\end{equation}
The derivative of $V$ with respect to $\bsM_{l:}$ is given by
\begin{equation}
\frac{\partial V}{\partial \bsM_{l:}} = - \left[\bLam_{l:} \odot \left( \bsY_{l:} -\bsM_{l:} \bsA \right)  \right] \bsA^T + \frac{1}{\epsilon^2}  \left(\bsM_{l:} - \widetilde{\bsM}_{l:}\right).
\end{equation}
The derivatives of $W$ with respect to $\bSig_{:l}$ are given by
\begin{eqnarray}
\frac{\partial W_1}{\partial \bSig_{:l}} & = &  - \frac{1}{2} \left[\frac{\left(\bsY_{l:}-\bsM_{l:} \bsA\right) \odot \left(\bsY_{l:}-\bsM_{l:} \bsA\right)}{\bOme_{l:} \odot\bOme_{l:} } \right]  \left(\bsA \odot \bsA\right)^T \nonumber \\
\frac{\partial W_2}{\partial \bSig_{rl}^2} & = & \frac{\partial W_2}{\partial \sigma_{rl}^2} = \frac{1}{\sigma_{rl}^2}, \; \forall r \in \left\lbrace 1,\cdots,R \right\rbrace \nonumber \\
\frac{\partial W_3}{\partial \bSig_{:l}} & = &
\frac{1}{2} \left[\bLam_{l:} \left(\bsA \odot \bsA \right)^T \right]
\nonumber \\
\label{eqt:dW_dSig}
\end{eqnarray}
The derivatives of $H$ with respect to $ \psi_{n}^2$ is given by
\begin{equation}
\frac{\partial T}{\partial \psi_{n}^2} =  \frac{\partial U_1}{\partial \psi_{n}^2}+\frac{\partial U_3}{\partial \psi_{n}^2}+\lambda
\end{equation}
with
\begin{eqnarray}
\frac{\partial U_1}{\partial \psi_{n}^2} & = &  -\frac{1}{2}
\sum_{l=1}^{L}  \frac{\left(\bsy_{n}-\bsM \bsa_{n}\right) \odot\left(\bsy_{n}-\bsM \bsa_{n}\right) }{\bOme_{:n} \odot\bOme_{:n} }
\nonumber \\
\frac{\partial U_3}{\partial \psi_{n}^2} & = &  -\frac{1}{2}
\sum_{l=1}^{L} {\bLam_{:n}}   \nonumber \\
\label{eqt:dUi_dpsi}
\end{eqnarray}
The derivative  of $P$ with respect to $\bsc_{rk}$ is given by
\begin{equation}
\frac{\partial P}{\partial \bsc_{rk}} =  \frac{\partial P_1}{\partial \bsc_{rk}}+\frac{\partial P_2}{\partial \bsc_{rk}}
\end{equation}
with
\begin{eqnarray}
\frac{\partial P_1}{\partial \bsc_{rk}} & = & \left(\gamma+1 \right) \sum_{n \in  \mathcal{I}_k}{  \left[
-  \Upsilon\left(\sum_{r'=1}^{R}{ c_{r'k}}\right) + \Upsilon \left(c_{rk} \right)
\right]  } \nonumber \\
\frac{\partial P_2}{\partial \bsc_{rk}} & = & \sum_{n \in  \mathcal{I}_k}{\left[\alpha - \log\left(a_{rn}\right)\right]}
 \nonumber \\
\label{eqt:dUi_dpsi}
\end{eqnarray}
where $\Upsilon$ denotes the polygamma function, i.e., the derivative of the log-gamma function.
 
%\section*{Acknowledgments}
%Nicolas Dobigeon is grateful to Yoann Altmann and Abderrahim Halimi,
%University of Toulouse, for sharing their unmixing algorithm MATLAB
%codes and for stimulating discussions.

%%\newpage
%\bibliographystyle{ieeetran}
%\bibliography{D:/Dropbox/strings_all_ref,D:/Dropbox/all_ref}

\newpage
\bibliographystyle{ieeetran}
\bibliography{biblio_all}

\end{document}